\documentclass[fleqn,usenatbib]{mnras}

\usepackage{newtxtext,newtxmath}
\usepackage{breqn}
\usepackage{enumerate}
\usepackage[T1]{fontenc}
\usepackage{ae,aecompl}

\usepackage{graphicx}	% Including figure files
\usepackage{amsmath}	% Advanced maths commands
\usepackage{xcolor}
\title[Thermal Scattering in Cloudy Exoplanets]{How Does Thermal Scattering Shape the Infrared Spectra of Cloudy Exoplanets? A Theoretical Framework and Consequences for Atmospheric Retrievals in the JWST era}

\author[Taylor et al.]{
Jake Taylor$^{1}$\thanks{E-mail: jake.taylor@physics.ox.ac.uk},
Vivien Parmentier$^{1}$,
Michael R. Line$^{2}$,
Elspeth K. H. Lee$^{3}$,
\newauthor
Patrick G. J. Irwin$^{1}$ \& Suzanne Aigrain$^{4}$
\\
$^{1}$Department of Physics (Atmospheric, Oceanic and Planetary Physics), University of Oxford, Parks Rd, Oxford, OX1 3PU, UK\\
$^{2}$School of Earth \& Space Exploration, Arizona State University, Tempe AZ 85287, USA \\
$^{3}$Center for Space and Habitability, University of Bern, Gesellschaftsstrasse 6, Bern, 3012, Switzerland \\
$^{4}$Department of Physics (Astrophysics), University of Oxford,
Denys Wilkinson Building, Keble Rd,
Oxford, OX1 3RH, UK 
}

\date{Accepted XXX. Received YYY; in original form ZZZ}

\pubyear{2021}

\begin{document}
\label{firstpage}
\pagerange{\pageref{firstpage}--\pageref{lastpage}}
\maketitle

% Abstract of the paper
\begin{abstract}
Observational studies of exoplanets are suggestive of a ubiquitous presence of clouds. The current modelling techniques used in emission to account for the clouds tend to require prior knowledge of the cloud condensing species and often do not consider the scattering effects of the cloud. We explore the effects that thermal scattering has on the emission spectra by modelling a suite of hot Jupiter atmospheres with varying cloud single-scattering albedos (SSAs) and temperature profiles. We examine cases ranging from simple isothermal conditions to more complex structures and physically driven cloud modelling. We show that scattering from nightside clouds would lead to brightness temperatures that are cooler than the real atmospheric temperature if scattering is unaccounted for. We show that scattering can produce spectral signatures in the emission spectrum even for isothermal atmospheres. We identify the retrieval degeneracies and biases that arise in the context of simulated JWST spectra when the scattering from the clouds dominates the spectral shape. Finally, we propose a novel method of fitting the SSA spectrum of the cloud in emission retrievals, using a technique that does not require any prior knowledge of the cloud chemical or physical properties. 

\end{abstract}

% Select between one and six entries from the list of approved keywords.
% Don't make up new ones.

%%%%%%%%%%%%%%%%%%%%%%%%%%%%%%%%%%%%%%%%%%%%%%%%%%

%%%%%%%%%%%%%%%%% BODY OF PAPER %%%%%%%%%%%%%%%%%%

\section{Introduction}

Clouds are present in all solar-system atmospheres and have been shown to be ubiquitous in exoplanet atmospheres~\citep{iyer2016characteristic,sing2016continuum,Wakeford2019}. They often hinder our ability to probe large regions of the atmosphere and are currently the main source of uncertainty when estimating the abundances of chemical species from exoplanet spectra \citep{Deming2013}, a situation that will still remain a challenge with the arrival of the James Webb Space Telescope (JWST)
~\citep{greene2016characterizing}.

Currently, most studies addressing the role of clouds in atmospheric retrievals have tackled the case of transmission spectra \citep{2014Lee,Barstow2016,2017Macdonald,wakeford2017hst,fisher2018retrieval,tsiaras2018population,Benneke2019,2020Carter}. They have shown that different cloud parameterisations lead to similar retrieved abundances~\citep{Mai2019,2020BarstowClouds}, but that neglecting the presence of clouds or neglecting their spatial variability could lead to incorrect retrieved abundances
~\citep{Line2016,2019Welbanks,2019Pinhas}.

Another way to sense exoplanet atmospheres is through their emission spectrum. Phase curves can be used to measure the emission spectrum of the planet at different phase angle, from the dayside to the nightside. Phase curves have been used to build temperature and cloud maps, showing the prevalence of clear and hot daysides \citep{kreidberg2014precise,Parmentier2016,stevenson2017spitzer,2020Irwin} and cloudy and cold nightsides \citep{Showman2009,Kataria2015,Parmentier2016,Helling2019,Beatty2019,Keating2019}. Furthermore, the radiative effect of nightside clouds have been shown to complicate the link between observable quantities, such as the phase curve amplitude and fundamental atmospheric properties such as the heat redistribution efficiency \citep{Parmentier2021}. 

Several cloud formation models of varying degrees of complexity have been used to investigate the formation of clouds and the diversity of their compositions in exoplanetary atmospheres, \citep[e.g.][]{2001Ackerman, 2003Woitke,Sharp2007, 2008Helling,2018Ohno, 2018Powell,Samra2020}. Most recently, \citet{Gao2020} use microphysical models to predict the most likely outcome of cloud formation in planetary atmospheres, concluding that for planets with a temperature greater than 800K the clouds would be silicate dominated. The impact of these various cloud compositions have been thoroughly explored in transmission \citep{wakeford2015transmission,pinhas2017signatures,Kitzmann2018,Mai2019}, and it has been found that the variation in cloud optical properties leads to a wide range in spectral shapes. 

The importance of clouds, particularly their scattering properties, in exoplanet emission spectra has been highlighted in the context of directly imaged planets and brown dwarfs \citep{2013Lee,2017Lavie,Burningham2017,2020Molliere}.  
However, the effects of scattering on the emission spectra of transiting exoplanets has been relatively unexplored; it was thought that the complicated effects of multiple scattering were negligible for the data quality. \citet{2014Barstow} used optimal estimation and a grid model search to study the effects of multiple scattering on the dayside reflectance spectrum of HD189733b, but found little evidence of strong reflection. \citet{deKok2011} explore the effects of multiple scattering using forward models and their findings suggest that the scattering of clouds in thermal emission can cause up to a 10\% difference in spectra compared to their cloud-free counterpart. Our work builds upon these by more deeply exploring the biases and degeneracies involving multiple scattering in the thermal infrared.

Exoplanet observations by JWST and ARIEL \citep{tinetti2018chemical} will improve our understanding of clouds by increasing the wavelength range of observations in the infrared compared to current HST capabilities. The phase curves planned as part of the JWST transiting exoplanets Early Release Science (ERS) program \citep{2018Bean} and the many more to come will allow for the study of night-side clouds, as well as the clouds at different phase angles, and thus reveal the cloud distribution in more detail.  \citet{Parmentier2016} predict that the day-side of hot Jupiters should be relatively cloud free. However, this may not be the case for cooler objects, with longer wavelength spectra accessible by JWST. Hence, when used to observe the night-side of hot Jupiters or to observe cooler planets, which are likely to sustain day-side clouds, the scattering effect will be prominent and might bias atmospheric parameter estimates. With this in mind, it is important that we begin to explore the effects of cloud droplet multiple scattering on thermal emission spectra over the broad wavelengths likely probed by JWST, as well as the corresponding biases and degeneracies that might arise within common parameter estimation frameworks.

An additional consideration is the way in cloud parameterisations are treated within Bayesian retrieval frameworks, usually relying upon \textit{a priori} knowledge of which condensate species are most prominent. For example, the retrievals conducted in \citet{Venot2020} assume\footnote{The emission retrievals with \textsc{TauREx} and \textsc{Pyrat Bay} used in this work did not account for multiple scattering in emission} the species of the cloud \textit{a priori}. Hence, they use fixed properties for the cloud such as its mean molecular weight and SSA. Given that such assumptions cannot generally be made \textit{a priori} when exploring the atmosphere of an unknown planet, it would be preferable to take a more agnostic approach, e.g.\ by using general parameterisations styles that can encompass a range of cloud compositions and configurations. Given that the scattering properties of clouds depend on the real and imaginary refractive index as well as on particle size, these three properties are frequently retrieved simultaneously for the solar system planets (which generally entail in situ high resolution, signal-to-noise spectra at multiple viewing angles) \citep{2015Irwin,2019Toledo,2020Braude}. This technique requires a large number of free parameters, which is problematic for typical exoplanet spectral coverage and precision (even for JWST quality). Instead, we propose a "middle ground" parameterization that fits for a simplified SSA wavelength dependence of the cloud species, as this property encompasses the effects of both the real and imaginary index spectra and the particle size. In this approach, we are agnostic to the chemical and physical properties of the cloud. 

This study aims to understand the role of clouds, and particularly thermal scattering, in shaping the emission spectra of exoplanets. This will help to anticipate the challenges that will arise when interpreting JWST quality spectra, and to provide mitigation strategies that are flexible and independent of the numerous cloud parameters that are both hard to predict and hard to constrain. In Section \ref{Section3} we describe our model set up, including how we parameterise the clouds. Section \ref{Section2} discusses why scattering in emission is important and how it effects the spectrum. Section \ref{Section4} explores a simplistic atmospheric set up, where the SSA for the cloud is independent of wavelength. In Section \ref{Section5} we then use more realistic optical properties for the clouds, based on a flexible three-point parameterisation of the SSA, which captures the main behaviour of the different cloud species.

\section{Numerical model setup}
\label{Section3}

In this section we outline the various models and techniques that will be used in this study. We specifically outline two different temperature-pressure (TP) parameterisations and three different cloud assumptions, and introduce a new cloud retrieval approach.

\subsection{NEMESIS}
We use the \textbf{N}on-linear optimal \textbf{E}stimator for \textbf{M}ultivariat\textbf{E} spectral analy\textbf{SIS} code \textsc{NEMESIS} \citep{irwin2008nemesis} to compute our model spectra. \textsc{NEMESIS} uses the correlated-$k$ approach \citep{lacis1991description} to model the spectra, which has been shown to be effective and accurate when compared to using the line-by-line or cross-section approaches \citep{garland2019effectively}. For this study the molecules used are: H$_2$O \citep{polyansky2018exomol} and CO \citep{li2015rovibrational}, which were formatted using the techniques presented in \citet{Chubb2020}. As we are modelling an atmosphere which is H$_2$-dominated we also modelled the H$_2$-H$_2$ and H$_2$-He collisionally induced absorption \citep{richard2012new}. \textsc{NEMESIS} calculates multiple scattering using the matrix operator method \citep{plass1973matrix}.

\subsection{Temperature-Pressure parameterisations}
\label{TP_param}
We consider two different parameterisations for the temperature-pressure (TP) profile in our study. The first is a simplistic three-parameter model, which allows us to fit for the gradient of the temperature profile. It takes the form:
\begin{equation}
\label{tp_1}
    T = T_0 \times \Big(\frac{P}{P_0}\Big)^{\alpha}
\end{equation}
where $T_0$, $P_0$ and $\alpha$ are free parameters. This parameterisation allows us to investigate how the gradient of the TP profile varies for the different model assumptions. We will only be using this TP parameterisation in Section \ref{non_iso}.
The second is the double gray analytic TP profile from  \citet{Parmentier2014}, which has five free parameters: $T_{\text{irr}}$, $\kappa_{\text{IR}}$, $\gamma_1$, $\gamma_2$ and $\alpha$. $T_{\text{irr}}$ is the irradiation temperature, $\kappa_{\text{IR}}$ is the infrared opacity, the parameters $\gamma_{1}$ and $\gamma_{2}$ are the ratios of the mean opacities in the two visible streams to the thermal stream: $\gamma_{1} = \kappa_{v1} / \kappa_{IR}$ and $\gamma_{2} = \kappa_{v2} / \kappa_{IR}$. The parameter $\alpha$ ranges between 0 and 1, and controls the weighting used between the two visible streams, $\kappa_{v1}$ and $\kappa_{v2}$. For each case we fix the internal temperature of the planet to be $T_{\text{int}}$ = 200\,K unless otherwise stated. For details of how this TP profile is implemented in \textsc{NEMESIS}, see \citet{Taylor2020}.

\subsection{Cloud parameterisation}
\label{cloud param}

Here we will briefly describe the various cloud parameterisations explored in this work. They all share the following similarities: The cloud is assumed to be a uniformly mixed gray opacity throughout the whole atmospheric column and the scattering phase function is assumed to be isotropic. The variations are as follows:
\begin{description}
    \item {\bf Purely absorbing case:} We set the SSA to 0.0. 
    This is analogous to a grey opacity cloud in other retrieval studies \citep[e.g.][]{line2016no}. 
    \item {\bf Constant SSA case:} We fit simultaneously for the cloud opacity and SSA, but take both to be constant with wavelength.
    \item {\bf Three-parameters case:} We model the SSA as a step function with one break point. The parameters are the SSA value before the break, the wavelength of the break, and the SSA value after the break. We call this retrieval style the "three-parameter" case as three free parameters are used to constrain the SSA. More break points could be added in future, but we adopted this prescription because it keeps the number of free parameters low, while adequately capturing the shape of the SSA spectrum of many cloud condensing species (Figure~\ref{fig:real_ssa}).
\end{description} 

\subsection{Parameter Estimation}
For the retrievals conducted in this study we have wrapped the forward-modelling component of \textsc{NEMESIS} in a Bayesian framework, using a nested sampling approach \citep{feroz2008multimodal,feroz2013importance}, which we implemented using \textsc{pymultinest} \citep{Buchner2014}. We use nested sampling due to its ability to compute both posterior distributions for the parameters, and the Bayesian evidence (or marginal likelihood), which can be used in model comparison and selection.

\section{How thermal scattering shapes the emission spectrum}
\label{Section2}
\subsection{Single Scattering Albedo and Emissivity}
\label{ssa_em}
Cloud particles can interact with photons by either absorption or scattering. The Single-Scattering Albedo (SSA) is the probability that a photon colliding with a particle is scattered rather than absorbed, and can be defined as:
\begin{equation}
    \omega_{\lambda}=\frac{Q_{\rm scat}(\lambda)}{Q_{\rm ext}(\lambda)}
\end{equation}
where the total extinction cross-section, $Q_{\rm ext}$ is the sum of the scattering cross-section $Q_{\rm scat}$ and the absorption cross-section $Q_{\rm abs}$. Additionally, the ability of particles to re-emit light is determined by their emissivity $\epsilon_\lambda$. In general, the source function in an atmosphere can be written as the combination of the scattering and the thermal emission terms:
\begin{equation}
    S_{\lambda}=\omega_\lambda J_\lambda+\epsilon_\lambda B_{\lambda}.
\end{equation}
where $B_\lambda$ is the Planck function and $J_\lambda$ is the first moment of the intensity. Kirchhoff's law states that at any particular wavelength, a particle is as good an emitter as it is an absorber. This means that the emissivity and the SSA must be linked by:
\begin{equation}
    \epsilon_\lambda+\omega_\lambda=1.
\end{equation}
The source function therefore only depends on the atmospheric emissivity (or only on the SSA) and can be used to solve the radiative transfer equations for a plane-parallel atmosphere at local thermodynamic equilibrium. The full solution for non-isothermal atmospheres can be found in~\citet{Rutten2003} and is reproduced in Appendix~\ref{general_rutten} for reference. In the following sections we discuss some simplified cases in more detail.

\subsection{Solution in the isothermal case}
\label{soln_iso_case}
An isothermal atmosphere is often thought to emit like a blackbody. However, when thermal scattering is present the source function is no longer a Planck function and the planetary flux becomes: 
\begin{equation}
    F_p = \frac{2\pi}{\sqrt{3}} \frac{2\sqrt{\epsilon_{\lambda}}}{1 + \sqrt{\epsilon_{\lambda}}} B_{\lambda},
    \label{eq::FLux}
\end{equation}
where the numerical prefactor depends on the choices of boundary conditions~\citep[see][Section 2.3.2 for a discussion]{Parmentier2014}. For unit emissivity, $\epsilon_\lambda = 1$, we recover the standard result, namely the planetary flux is equal to the Planck function. For smaller emissivity, however, the outgoing planetary flux is smaller than the blackbody cases, and vanishes for a purely scattering atmosphere.

\begin{figure}
    \centering
    \includegraphics[width=0.45\textwidth]{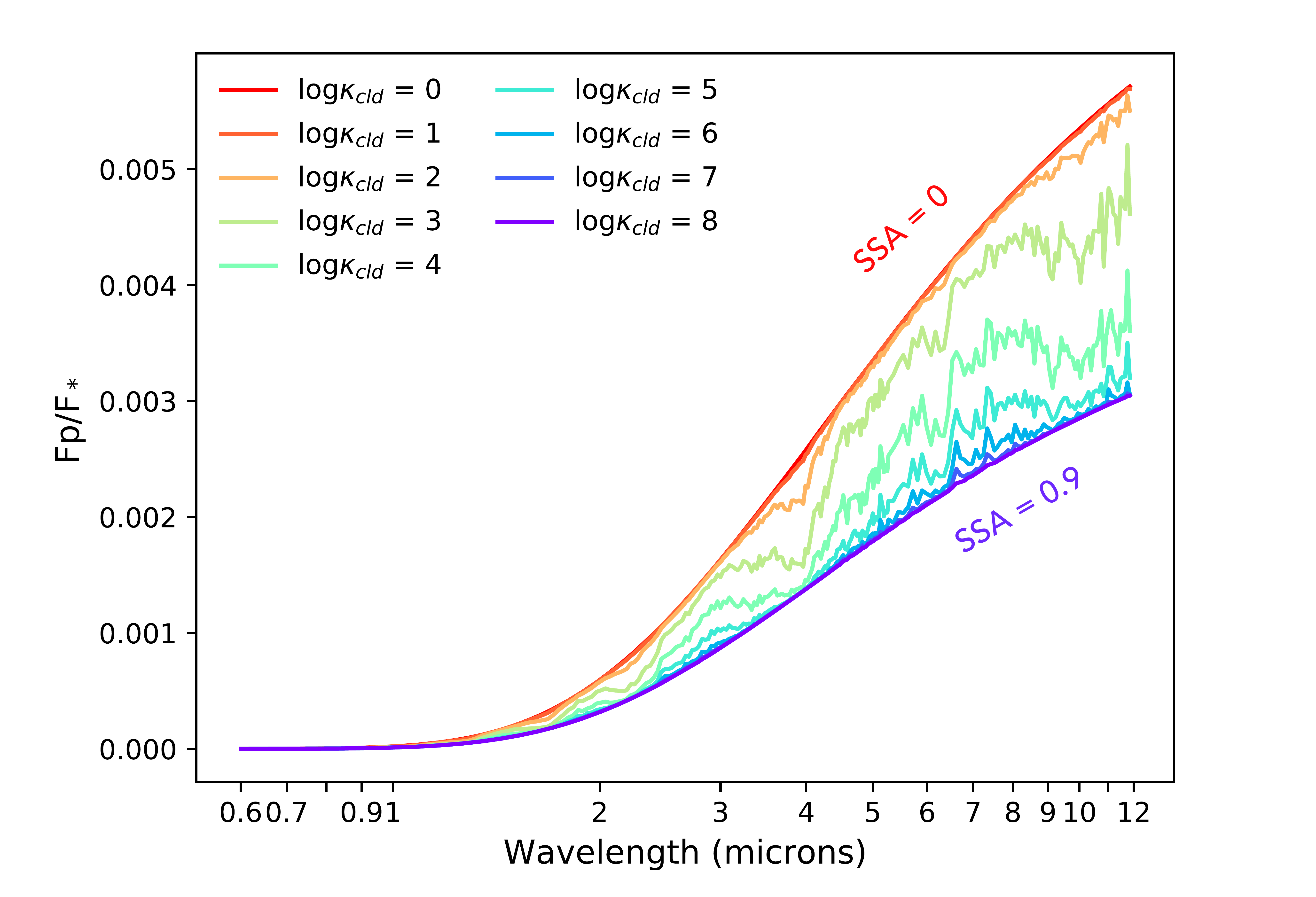}
    \caption{Model atmospheres varying from cloud free ($\text{SSA} = 0$, smooth red line) to completely dominated by the opacity of the cloud ($\text{SSA}=0.9$, smooth purple line). The thinner lines containing features, which are due to H$_2$O in the atmosphere, correspond to isothermal atmospheres with a temperature of 1400\,K and a cloud with a SSA of 0.9, whose opacity increases by an order of magnitude between each line, from red to purple. We transition from a molecule opacity dominated regime to a cloud opacity dominated regime. It can be seen that the opacity of the cloud results in different temperature layers being probed. }
    \label{fig:diff_opac}
\end{figure}

When the emissivity is constant with wavelength, the resulting planetary spectrum is a diluted blackbody. This is demonstrated in Figure~\ref{fig:diff_opac}, where the purple curve corresponds to a model entirely dominated by a cloud opacity, with SSA $\omega=0.9$, and hence emissivity $\epsilon=0.1$. 

\begin{figure*}
    \centering
    \includegraphics[width=\textwidth]{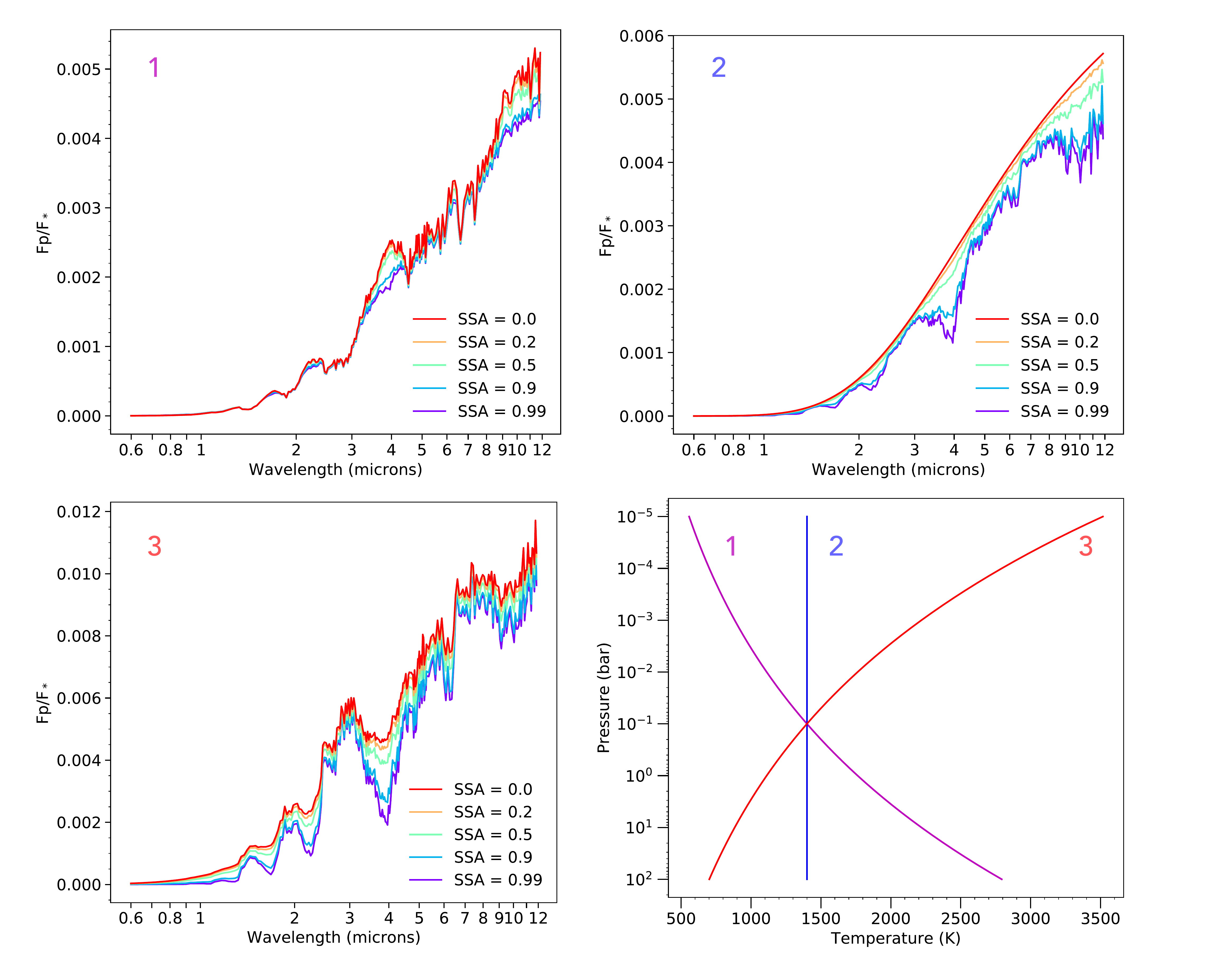}
    \caption{Impact of the SSA on the emission spectra for different TP profiles. The TP profiles corresponding to cases 1, 2 and 3 (top left, top right and bottom left plot, respectively) are shown in the bottom right plot. All models were generated using $\log(\kappa_{\rm cld}) = 3$ and with roughly solar H$_2$O and CO abundances. The non-isothermal TP profiles were generated using the 3-parameter model discussed in Section~\protect{\ref{TP_param}}
    (see Eq.~\protect{\ref{tp_1}}), with $\alpha = 0.1$ and $-0.1$ for cases 1 and 3, respectively.}
    \label{fig:compare_ssa}
\end{figure*}

In the general case, emissivity should depend on wavelength for two reasons. First, the cloud emissivity itself depends on wavelength (see Section~\ref{ssa_em} and Fig.~\ref{fig:compare_ssa}). Second, the local emissivity is the weighted sum of the cloud and gas emissivities. As we take the latter to be unity in this work, the local emissivity amounts to:
\begin{equation}
    \epsilon_{\lambda}=\frac{\epsilon_{\rm cld} \, \kappa_{\rm cld}+\kappa_{\rm gas}}{\kappa_{\rm cld}+\kappa_{\rm gas}},
\end{equation}
where $\kappa_{\rm cld}$ and $\kappa_{\rm gas}$ are the cloud and gas opacities, respectively. 

This equation shows that, for a constant cloud opacity and emissivity, the local emissivity is a monotonically decreasing function of the molecular opacity. The dilution factor before the Planck function in Equation \ref{eq::FLux} is then a monotonically-decreasing function of the molecular opacities. As a consequence, inside molecular absorption bands the emissivity will be high, the dilution factor will be small and the planetary flux will be large, whereas it is the opposite outside molecular bands. In summary, \emph{an isothermal atmosphere with a cloud that has a constant opacity and constant emissivity will mimic molecular emission features}. 

Figure~\ref{fig:diff_opac} demonstrates this behaviour for an isothermal atmosphere with H$_2$O and CO in solar-composition proportions and a constant opacity, constant emissivity cloud. If the cloud opacity is much smaller than the gas opacity, the spectrum follows a Planck function as expected. As the cloud opacity increases, it becomes comparable to the gas opacities, and emission bands of H$_2$O and CO start to appear. For very large cloud opacity, the emissivity is entirely dominated by the cloud contribution and the spectrum reverts to that of a blackbody, albeit diluted by a constant factor. 

\subsection{Non-isothermal case}
\label{non iso case}
In the non-isothermal case (see derivation in Appendix \ref{general_rutten}) the situation is more complex and the planetary flux takes the form: 
\begin{equation}
        F_p=\frac{2\pi}{\sqrt{3}}\left(\frac{2\sqrt{\epsilon_{\lambda}}}{1+\sqrt{\epsilon_\lambda}}\right)B_{\lambda}\left(\tau_\lambda=\frac{1}{\sqrt{3\epsilon_\lambda}}\right).
\end{equation}
Scattering therefore affects the planetary flux in two ways. First, as for the isothermal case, it multiplies the planetary flux by a dilution factor, which is a monotonically decreasing function of the gaseous opacity. Second, it increases the effective depth of the photosphere by a factor $1/\sqrt{\epsilon_\lambda}$. 

For an atmosphere with a temperature that increases with increasing altitude, the flux is generally smaller outside molecular bands. However, that is where the emissivity is low, hence the photosphere probes deeper, colder layers and the flux is reduced. Similarly, the dilution factor is small when the emissivity is low. Thus, for an atmosphere with a temperature inversion and emission features, both effects act jointly to reduce the planetary flux more strongly outside molecular bands than within them, hence increasing the relative contrast of the emission features. 

For an atmosphere with a temperature that decreases with increasing altitude, the flux is generally larger outside molecular bands. However, that is a region where the emissivity is low, hence the photosphere probes deeper, hotter layers. We see an opposite effect with the dilution factor, as it is small when the emissivity is low. Therefore, for an atmosphere without a temperature inversion, the two effects of the scattering act in opposition. Whether the amplitude of molecular features is increased or decreased by scattering depends on the strength of the temperature gradient. 

These different behaviours are shown in Figure~\ref{fig:compare_ssa}. In the first panel, we show the case of a non-inverted atmosphere with a temperature gradient that is weak enough that the effect of the dilution factor dominates, and the amplitude of the molecular features decreases with increasing SSA. In the second panel, we show the case of an isothermal atmosphere, where emission features appear as the SSA increases. Finally, the third panel show that case of an inverted atmosphere, where the strength of the emission features increases with increasing SSA.  

\subsection{Expected degeneracies}
\label{degen}
We have shown that thermal scattering can change the shape of the spectral features in a planet atmosphere and even create emission features in the isothermal case. This will naturally lead to degeneracies between parameters: the spectra produced by atmospheres with scattering can look very much like those of atmospheres without scattering, but with a different set of physical properties. 

The most obvious degeneracy arises when comparing the brightness temperature of a planet to the actual atmospheric temperature. Indeed, for an isothermal atmosphere with a constant, non-zero emissivity, Eq.~(\ref{eq::FLux}) shows that the spectrum is a diluted blackbody. If we observe the atmosphere at long wavelengths, where the Planck function is linear in temperature, the brightness temperature measured in a given bandpass would become: 
\begin{equation}
\label{t_bright}
    T_{\rm bright}=\frac{2\sqrt{\epsilon}}{1+\sqrt{\epsilon}}~T_{\rm atm}. 
\end{equation}
For example, measuring the brightness temperature from the spectrum produced by an atmosphere with an emissivity of $\epsilon=0.5$ would lead to a retrieved brightness temperature roughly $20\%$ smaller than the actual atmospheric temperature. With an emissivity of $\epsilon=0.1$, the brightness temperature is roughly $50\%$ smaller than the true atmospheric temperature.  This has far reaching implications for the night-side of hot Jupiters, which have been shown to have cooler than expected temperatures \citep{Beatty2019,Keating2019,Parmentier2021}. These results could be explained by the presence of low emissivity, high single-scattering clouds, such as pure silicate clouds, which would lead to an apparent brightness temperature much smaller than the actual atmospheric temperature. 

The second degeneracy arise from the dilution factor itself. \cite{Taylor2020} showed, in the context of a planet with an inhomogeneous thermal structure, that such a dilution factor can be mimicked by the spurious addition of molecules such as methane or ammonia. We therefore expect that the lack of scattering in retrieval models could lead to the spurious detection of molecules in cloudy atmospheres. 

The third degeneracy we can expect is linked to the spectral variation of the atmospheric emissivity, which could affect either the inferred strength and sign of the temperature gradient, or the inferred molecular abundances. As highlighted in Fig.~\ref{fig:compare_ssa}, the wavelength dependence of the emissivity generally reduces absorption features and increases emission features. This can be compensated in at least three different ways by a non-scattering retrieval. First, the thermal structure could be biased towards more positive temperature gradients. Second, the molecular abundances could be biased towards lower values for non-inverted atmospheres, so that the increased pressure broadening of the lines reduces the spectral features. Conversely, the abundances could be biased towards larger values for inverted atmospheres, so that the reduction of the pressure broadening of the lines enhances the size of the features. Finally, the change in the strength of molecular features could be compensated by the addition of a gradient in the molecular abundances. As shown by
~\citet{Parmentier2018}, spectral features are increased if abundances increase with height and decreased if abundances decrease with height. A more thorough discussion of these degeneracies, based on an analytical derivation can be found in Appendix ~\ref{Appendix B}

\section{Results from constant cloud single-scattering albedo case}
\label{Section4}

In this section we start with a simplistic model to understand the biases and degeneracies that thermal scattering can induce when extracting information from the emission spectra of exoplanets. For this we focus on the addition of a cloud with a constant opacity and constant emissivity. We first explore the simplest case of an isothermal atmosphere and then move to the non-isothermal case.

\subsection{Isothermal atmosphere}
\label{iso const}

As seen in Section \ref{soln_iso_case} we expect a cloud with a constant SSA in an isothermal solar-composition atmosphere to produce features that mimic emission that could be interpreted as the presence of an inverted thermal structure. We now want to quantify this in a retrieval framework.

We base our physical parameters of the system around the canonical exoplanet WASP-43b \citep{Hellier2011} which have been refined by \citet{gillon2012trappist}. We present the physical parameters and atmospheric parameters used in Table \ref{tab:model_params}.

\begin{table}
   \centering
\begin{tabular}{l|l}
    \hline
    Parameter      & Value   \\
    \hline
    log(H$_2$O)     & -3 \\
    log(CO)     & -3 \\
    Temp [K]     & 1400 \\
    SSA & 0.9 \\
    log(Opacity) & 4

%    \hline
\end{tabular}
\caption{Atmospheric and cloud properties used to create the isothermal model atmosphere.}
    \label{tab:model_params}
\end{table}
We used \textsc{NEMESIS} to generate a forward-modelled spectrum given the atmospheric parameters in Table \ref{tab:model_params} and then we used \textsc{PandExo} to generate the wavelength grid and the error bars for a single eclipse observation of WASP-43b. We consider the observing modes by NIRSpec Prism and MIRI LRS. We generate these observations considering a resolution of R = 50. We do not employ a random noise instance, but rather assume the data points at the exact model value in order to mitigate biases due to random noise instances \citep{Feng2018,Mai2019,Changeat2019}

In order to estimate which wavelengths are most sensitive to the effects of thermal scattering we focus on three different possible observing strategies:
\begin{enumerate}[1.]
    \item[1.] A single eclipse observation using the NIRSpec Prism observing mode. Wavelength coverage: 0.6 - 5.3 $\mu$m
    \item[2.] A single eclipse observation using the MIRI LRS observation mode. Wavelength coverage: 5.0 - 12.0 $\mu$m
    \item[3.] Combining both NIRSpec Prism and MIRI LRS observations to investigate the complete wavelength coverage of JWST (0.6 - 12.0 $\mu$m).
\end{enumerate}

We then perform retrievals of temperature and composition assuming the \citet{Parmentier2014} pressure-temperature profile from Section \ref{TP_param} and an atmosphere with H$_2$O and CO gaseous absorption. Additionally, each of our three retrievals treats the clouds with an increasing complexity: 
\begin{enumerate}[1.]
    \item A cloud-free model, shown in blue in upcoming figures.
    \item A pure absorbing cloud with constant opacity as described in Section \ref{cloud param}  This model is shown in red in upcoming figures. 
    \item A constant single-scattering albedo (SSA) cloud, with a constant emissivity and a constant opacity as described in Section \ref{cloud param}. This model is shown in purple in upcoming figures.
\end{enumerate}

In order to quantify the fit each model, we calculate the Bayesian evidence and convert the result to a sigma significance following the same method as \citet{Taylor2020}. We compare the cloud free and the pure absorbing cloud model to the scattering cloud model, as the scattering cloud model is representative of the "true" model, and is hence the reference model for the model comparison analysis. In this comparison, the larger the value of the sigma significance, the more rejected the model is. 

\subsubsection{Observing scenario 1: NIRSpec Prism (0.6 - 5.3 $\mu$m)}

\label{iso nirspec}

The wavelengths observed with NIRSpec Prism cover a broad array of molecular species encompassing the dominant nitrogen, oxygen, and carbon reservoirs. Hence, NIRSpec is one of the instruments which will be highly used by the exoplanet community \citep{stevenson2016transiting,2018Bean}. The NIRSpec Prism is the lowest resolution instrument at around R $\sim$ 100, but with the broadest simultaneous wavelength coverage.

\begin{figure*}
    \centering
    \includegraphics[width=\textwidth]{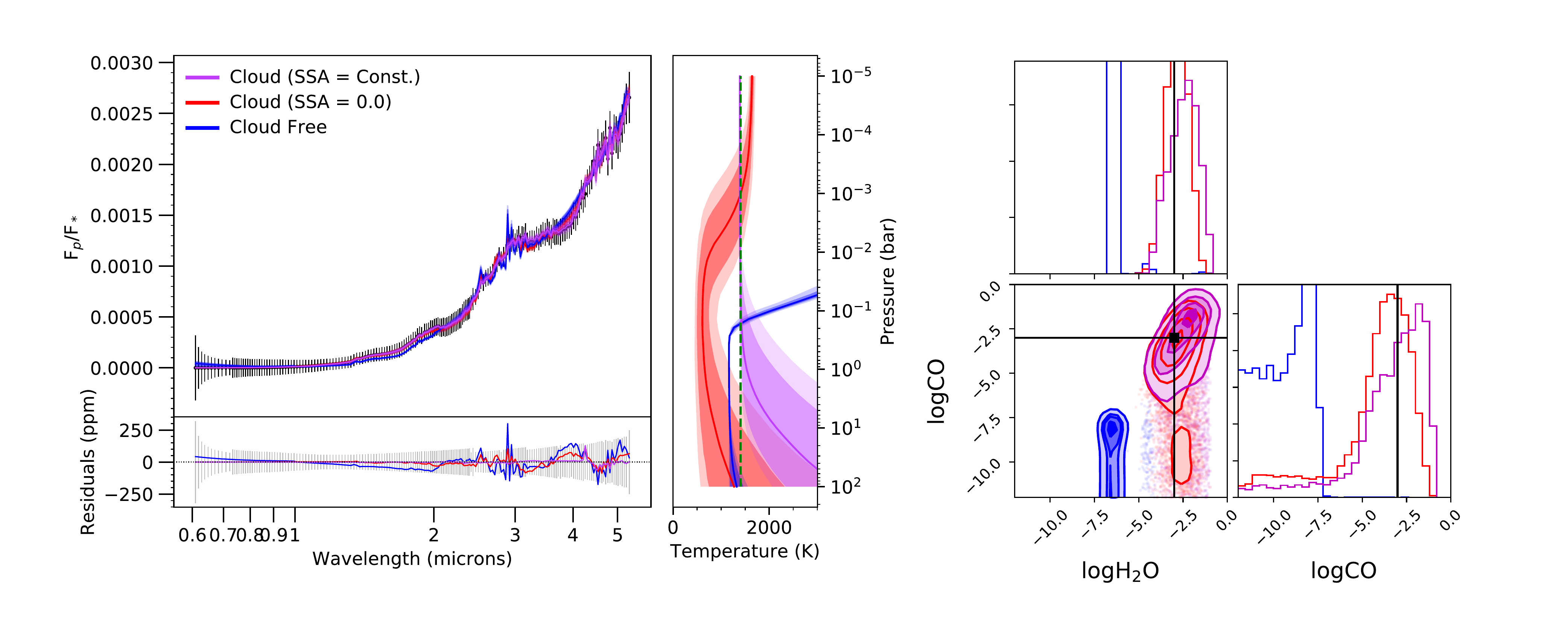}
    \caption{\textbf{Left}: Best-fitting spectra, temperature profiles and retrieved chemistry for three different retrieval models applied to synthetic NIRSpec Prism observations generated from planet with an isothermal TP profile and cloud opacity of $10^4$. For each model we retrieve the TP structure with the analytical TP profile described in Section \ref{TP_param}. In purple we retrieve on the synthetic data with the constant SSA cloud model. In red we fit using the pure absorbing cloud model. In blue we present the fit using a cloud-free model. We present the retrieved temperature-profiles with the true profile represented by a dashed green line. We present the residuals of the fit with grey shading indicating the error in ppm on the measurement. \textbf{Right}: we show the retrieved chemistry for each case with the true value in black.}
    \label{fig:NIRSpec_Iso_Pandexo}
\end{figure*}

Fig \ref{fig:NIRSpec_Iso_Pandexo} summarizes the retrieval results (fits, TP profile, and abundance constraints) under the 3 different retrieval model parameterisations. When calculating the sigma significance of excluding the cloud free and the pure absorbing cloud model we find that they are excluded at 8.3-$\sigma$ and 4.8-$\sigma$ respectively. Both the pure absorbing cloud and the cloud-free models require a temperature inversion to be able to fit to the data. However, when considering the scattering and retrieving for the SSA using the constant SSA cloud model, we are able to retrieve the correct profile, as we should given this retrieval forward model is identical to the input forward model.

Interestingly, the non-scattering cloud forward model is able to retrieve the CO and H$_2$O abundances with a similar precision and accuracy as the "true" scattering forward model. Despite this success, however, we would incorrectly retrieve an inverted thermal profile. This suggests, in this scenario, that the scattering-TP profile degeneracy is largely decoupled from the abundances. The cloud-free model is not able to retrieve the correct abundances. In fact the retrieved abundances for H$_2$O and CO are so low that it appears the model is using the H$_2$-H$_2$ collision induced absorption to fit to the synthetic data.

\subsubsection{Observing scenario 2: MIRI LRS (5.0 - 12.0 $\mu$m)}
\label{iso miri}
\begin{figure*}
    \centering
    \includegraphics[width=\textwidth]{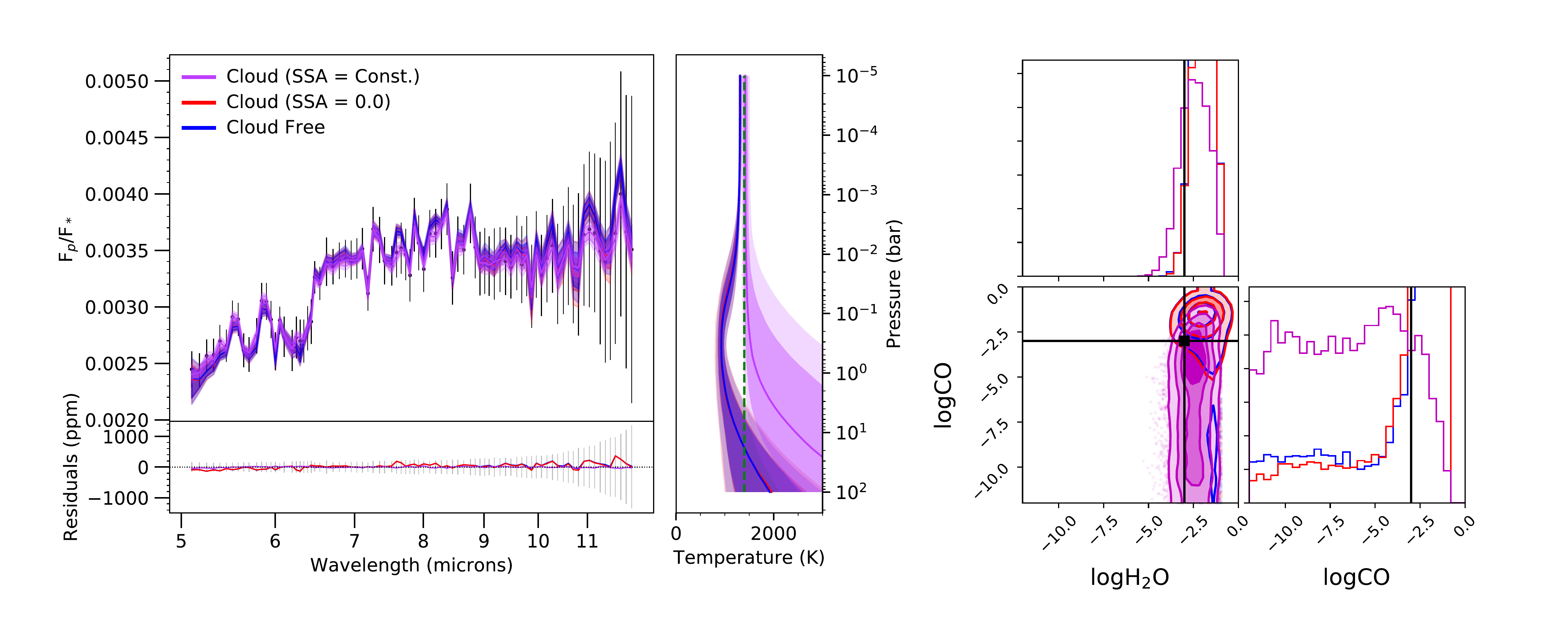}
    \caption{Same simulations as Figure \ref{fig:NIRSpec_Iso_Pandexo}, but for the MIRI LRS case.}
    \label{fig:MIRI_Iso_Pandexo}
\end{figure*}

The 5.0 - 12.0 $\mu$m spectral range contains the strongest H$_2$O features and is crucial in studying the clouds in exoplanet atmospheres as it covers such cloud features as a silicate `bump' \citep{Cushing2006,wakeford2015transmission}. Therefore, we have decided to explore its ability to recover the atmospheric properties from model observations generated using a simple scattering cloud. Fig \ref{fig:MIRI_Iso_Pandexo} summarizes the retrieval constraints (similar to Fig \ref{fig:NIRSpec_Iso_Pandexo}) achieved under the MIRI observational setup. 

The non-scattering cloud and cloud free retrievals produce nearly identical fits to the simulated data. This is due to the pure absorbing cloud model fitting for a near zero opacity which is consistent with a cloud-free atmosphere. When compared to the scattering cloud model, they can be excluded with a sigma significance of 4.3-$\sigma$ and 4.4-$\sigma$ respectively. Both these cases produce a slightly inverted TP profile. The residuals show that the fits are good given the error on the observations, and by not considering the effects of scattering, we could consider this atmosphere to be cloud-free with a temperature inversion. We also see that the retrieved TP profiles for these models are different from the TP profiles retrieved when studying the short wavelength case. The constant SSA cloud model fits best to the observations, with the TP profile converging to the true isothermal input. 

The constant SSA model is able to retrieve the abundance of H$_2$O, but not CO. This makes sense as there are no CO features in this wavelength region. The cloud-free and pure absorbing cloud model slightly overestimates the abundance of H$_2$O and find an upper limit on the CO which is higher than the input.

\subsubsection{Observing scenario 3: Full Wavelength Range}
\label{iso all}
\begin{figure*}
    \centering
    \includegraphics[width=\textwidth]{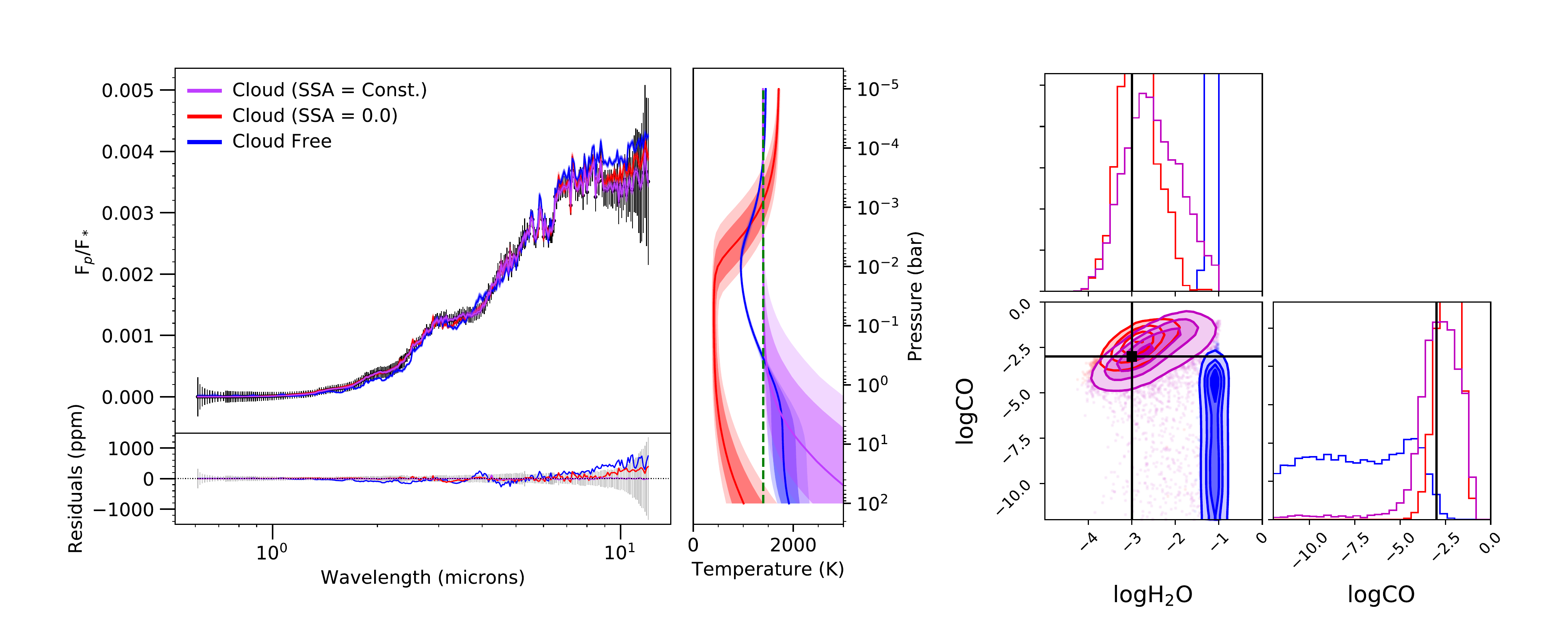}
    \caption{Same simulations as Figure \ref{fig:NIRSpec_Iso_Pandexo}, but for the combined NIRSpec Prism and MIRI LRS case. We present the full posterior distribution for the scattering cloud retrieval in Fig. \ref{fig:iso_cloud_posterior}.}
    \label{fig:ALL_Iso_Pandexo}
\end{figure*}

Figure \ref{fig:ALL_Iso_Pandexo} summarizes the retrieved constraints when combining the NIRSpec and MIRI modes for full 0.6 - 12 $\mu$m coverage.

By combining the observations the fit to the data is now worse for the cloud-free model, which now can be excluded with a sigma significance of 14.8-$\sigma$, with the most extreme deviations over the MIRI LRS wavelength coverage. It can be seen to have a slight temperature inversion and a retrieved H$_2$O abundance that is 2 orders of magnitude larger than the input value. We also find that we are unable to detect CO, despite it being present in the forward model.

We find that for the cloud-free models, the retrieved TP profiles for the short (0.5 - 5.3 $\mu$m) and long (5.0 - 12.0 $\mu$m) wavelength ranges are extremely different and mutually exclusive. As a consequence, when fitting the full wavelength range the cloud-free model is unable to provide a good fit to the synthetically generated observed data.

Interestingly, similar to the NIRSpec only case, the non-scattering cloud model produces abundance constraints reasonably consistent with the truth (and the scattering cloud constraints). The non-scattering cloud model can be excluded with a sigma significance of 7.1-$\sigma$. Hence, if we did not compare with a scattering cloud model, we could be tricked into believing that this model is the best fitting model and also conclude that this planet has a temperature inversion, which is not, in fact, the case and not what we would conclude using the constant SSA model.

\subsection{Non-isothermal atmosphere}
\label{non_iso}

We now explore the retrieval biases expected in a more realistic case of a planet with a temperature that decreases with decreasing pressure. The set up is the same as in Section \ref{iso const}, but this time the TP profile used is that described by three-parameter model in Section \ref{TP_param}, with $T_0$ = 1400 K, $P_0$ = 0.1 bars and $\alpha$ = 0.1. We also reduced the log$\kappa_{(\text{cld})}$ to be 3. Similar to Section \ref{iso const} we consider the same three observing cases.

We then re-retrieve on each data set with the three models described in Section \ref{iso const}. We use the three-parameter temperature profile for the non-isothermal case as we wanted to investigate the retrieved temperature-pressure slope to see if it is consistent with the analytical approach in Section \ref{non iso case}.

\subsubsection{Observing scenario 1: NIRSpec Prism}

\begin{figure*}
    \centering
    \includegraphics[width=\textwidth]{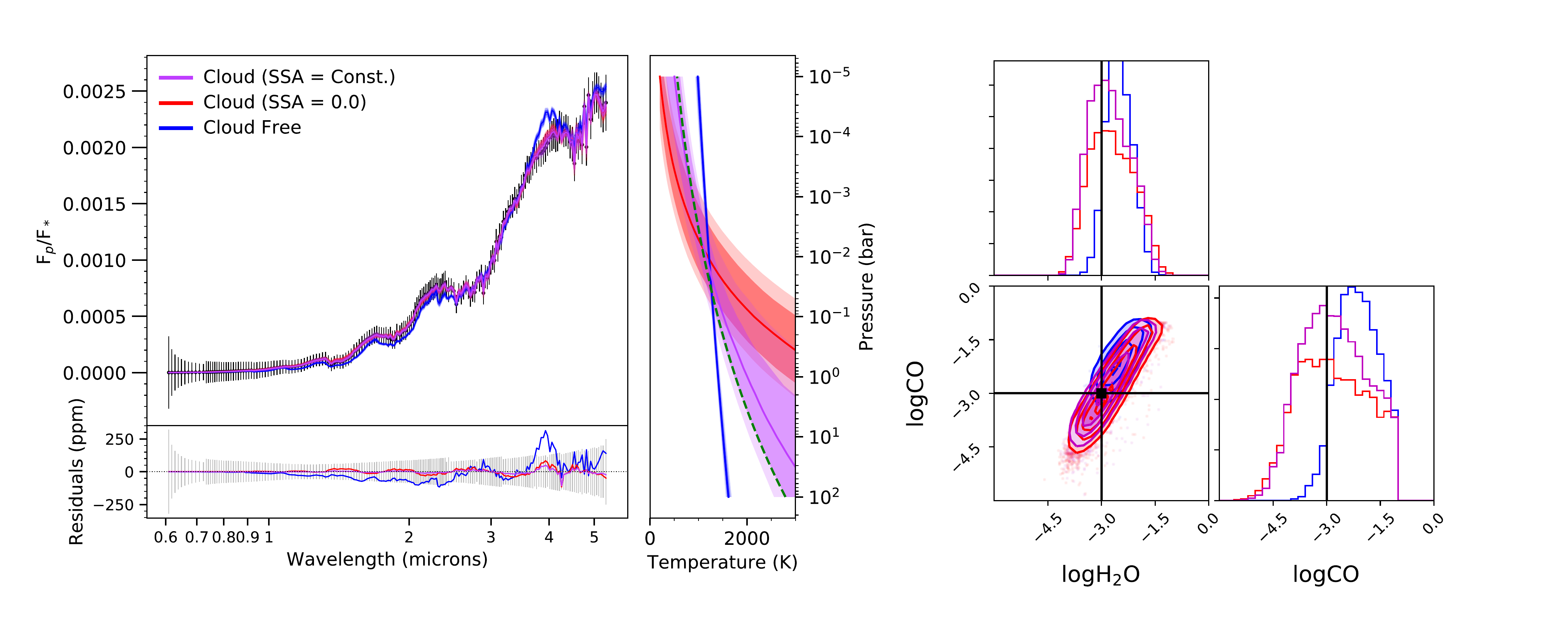}
    \caption{\textbf{Left:} Best-fitting spectra, temperature profiles and retrieved chemistry for three different retrieval models applied to synthetic NIRSpec Prism observations generated from a planet with three-parameter TP profile and cloud opacity of $10^3$.
    In purple we retrieve on the synthetic data with the 3-parameter TP profile described in Section \ref{TP_param} and model which allows for the single-scattering albedo to be fit as a constant value at all wavelengths. In red we retrieve using a similar model to the one shown in purple. However, we fix the single-scattering albedo to be zero, effectively removing the scattering component of the cloud and making the cloud grey. In blue we present a model that has no cloud. We present the retrieved temperature-profiles with the true profile represented by a dashed green line. We present the residuals of the fit with grey shading indicating the error in ppm on the measurement. \textbf{Right:} The posterior distributions of the retrieved chemistry with the colours corresponding to the models in the left plot. The black line represents the true value.}
    \label{fig:NIRSpec_Non_Iso_Pandexo}
\end{figure*}

We first investigate the three retrieval forward models on the NIRSpec Prism data, just as in Section \ref{iso nirspec}. Fig.  \ref{fig:NIRSpec_Non_Iso_Pandexo} summarises these results.

The residuals show that both the constant SSA model and pure absorbing cloud model have similar fits to the data and that the cloud-free model deviates the most, with significant deviation at around 4 $\mu$m, the region where CO should be present. The cloud free and pure absorbing cloud models have an excluded sigma significance of 9.3-$\sigma$ and 2.4-$\sigma$ respectively, suggesting that the cloud free model is firmly rejected by the data but the non-scattering cloud model cannot be rejected. For the cloud-free model, we retrieve a steeper TP profile compared to the input, as anticipated from the analytic arguments. Only the constant SSA model can retrieve the input TP profile, with the pure absorbing cloud model retrieving a larger temperature at higher pressures. 

All models are able to retrieve the correct composition to within 2-$\sigma$, with nearly consistent posteriors. The pure absorbing cloud model is the least accurate and has the widest posterior distribution.

\subsubsection{Observing scenario 2: MIRI LRS}

\begin{figure*}
    \centering
    \includegraphics[width=\textwidth]{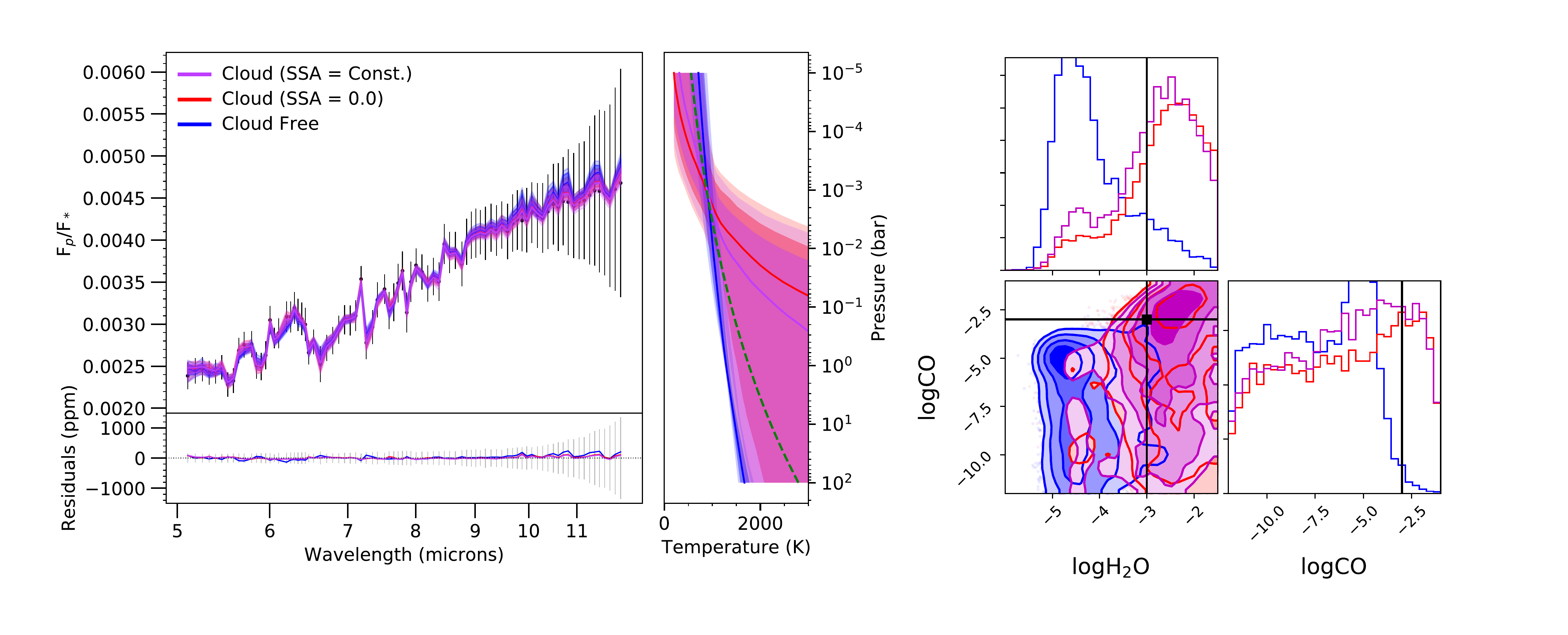}
    \caption{Same as Fig \ref{fig:NIRSpec_Non_Iso_Pandexo} but for the wavelength grid and errorbars of MIRI LRS.}
    \label{fig:MIRI_Non_Iso_Pandexo}
\end{figure*}

Fig \ref{fig:MIRI_Non_Iso_Pandexo} summarises the retrieval results under the MIRI LRS observational setup.

The residuals show that the cloud-free fit is not as good as either the pure absorbing cloud or SSA cloud models. The cloud free and pure absorbing cloud models are both considered consistent with the data (e.g., neither is rejected with cloud free at 2.3-$\sigma$, and purely absorbing at < 1-$\sigma$).  The same trends that can be seen in the NIRSpec Prism retrieved TP profiles are also seen with MIRI LRS. However, the profiles are less constrained. 

The abundance posteriors are much wider than those obtained using NIRSpec Prism, with the cloud-free model not being able to retrieved the correct chemistry. The pure absorbing cloud and the constant SSA cloud model are able to retrieve the H$_2$O abundance with a posterior distribution which spans 5 orders of magnitude. This is a result of the large errors on the data, the data quality is not good enough to distinguish from the different models or say anything meaningful about the abundances. 

\subsubsection{Observing scenario 3: Full Wavelength Range}

\begin{figure*}
    \centering
    \includegraphics[width=\textwidth]{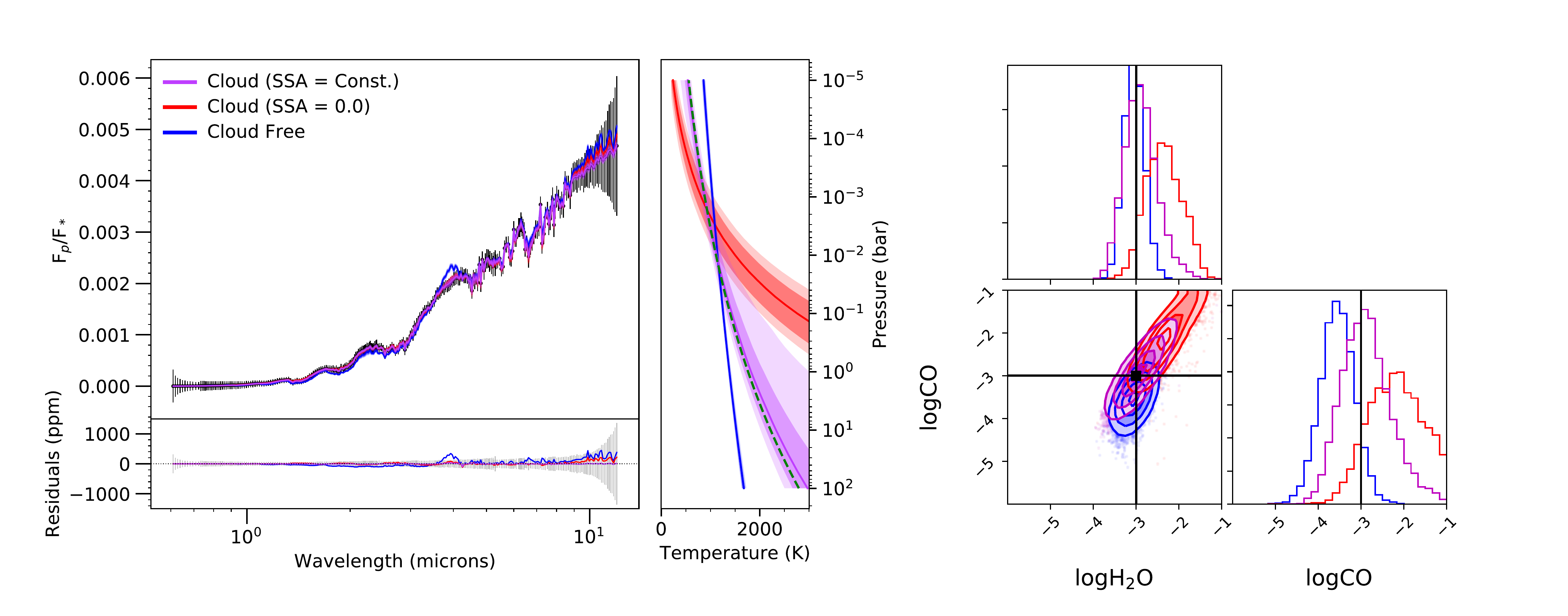}
    \caption{Same as Fig \ref{fig:NIRSpec_Non_Iso_Pandexo} but for the wavelength grid and errorbars of NIRSpec Prism and MIRI LRS.}
    \label{fig:ALL_Non_Iso_Pandexo}
\end{figure*}

Fig \ref{fig:ALL_Non_Iso_Pandexo} shows the constraints under the three retrieval forward model assumptions given the NIRSpec Prism + MIRI LRS "full" wavelength coverage observational scenario. The cloud free and pure absorbing cloud models have an excluded sigma significance of 11.2-$\sigma$ and 3.1-$\sigma$ respectively, this suggests that the cloud free model is firmly rejected, however the pure absorbing cloud is only marginally rejected, with the rejection limit being at 3-$\sigma$.

Interestingly, the cloud-free and the constant SSA cloud cases retrieve relatively similar H$_2$O abundances. However, the cloud-free model constrains a lower value of the CO abundance by around 2-$\sigma$ compared to the input value. The pure absorbing cloud model constrains higher values of both the H$_2$O and CO abundances by around 3-$\sigma$ compared to the input model value. This suggests that the broader wavelength coverage introduces more bias into the retrieved abundances when using the incorrect cloud model. 

We find that the broad wavelength coverage results in higher precision constraints on the TP profile when using the correct forward model. 

\subsection{Inverted Thermal Profile}
We now explore whether the biases exist when the atmosphere has an inverted thermal profile. We use the same setup from Section \ref{non_iso}, this time setting the $\alpha$ parameter to be $\alpha$ = -0.1 when generating our synthetic spectra, this results in an atmosphere with a thermal structure where the temperature increases with height. We then re-retrieve on this spectrum using the three models described in Section \ref{iso const}.

\begin{figure*}
    \centering
    \includegraphics[width=\textwidth]{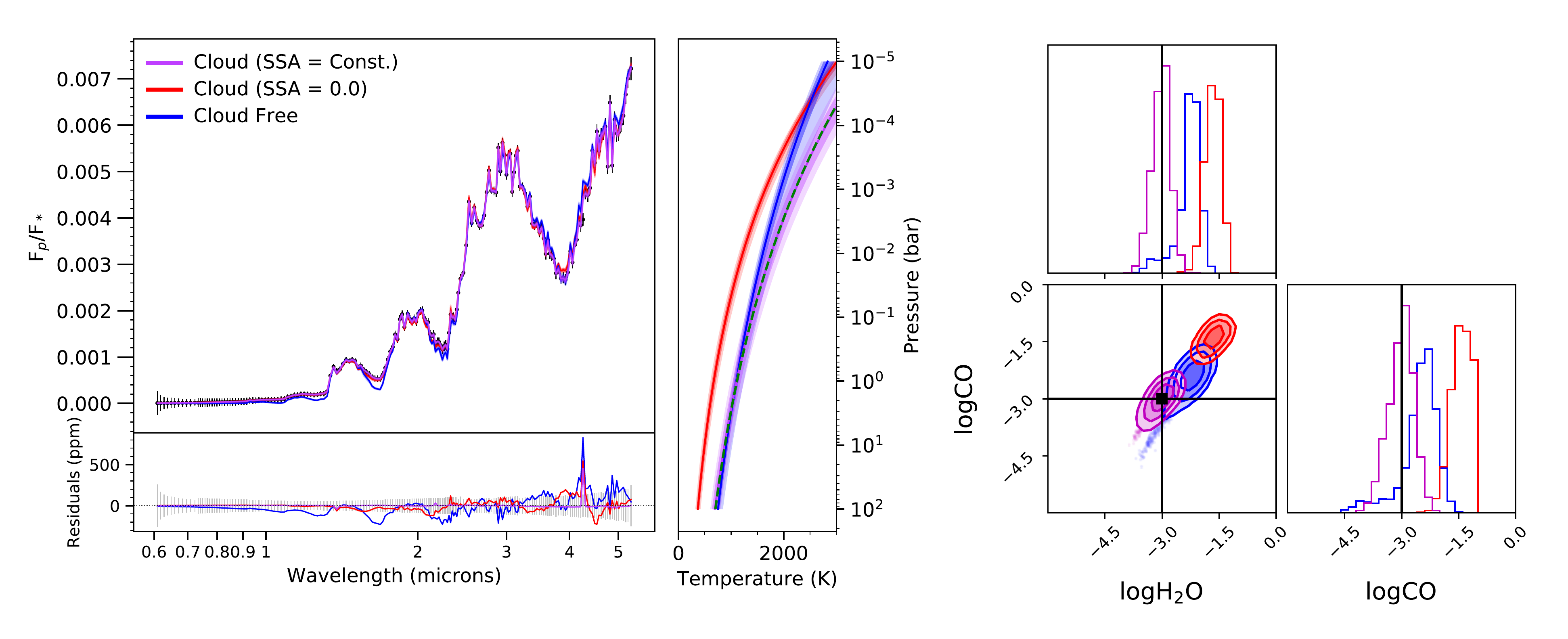}
    \caption{Same as Fig \ref{fig:NIRSpec_Non_Iso_Pandexo} but for the data generated using an inverted thermal profile.}
    \label{fig:NIRSpec_Inv}
\end{figure*}

For this case, we only consider the observing scenario of NIRSpec Prism. Fig. \ref{fig:NIRSpec_Inv} summaries these results. It can be seen the fit of the scattering cloud and the pure grey cloud models are consistent within the errorbars, however, the cloud free model shows the largest deviation. The cloud free and pure absorbing cloud models have an excluded sigma significance of 15.4-$\sigma$ and 8.0-$\sigma$ respectively, this suggests that both the cloud free and pure absorbing cloud models are firmly rejected. Only the scattering model is able to retrieve the input thermal structure and chemistry.

\subsection{Summary}

In summary, we demonstrated that failure to account for scattering can result in biased results in both the abundance inferences and in the derived thermal structures, though a greater bias appears in the temperature profiles. For the simple isothermal, scattering atmosphere,  the interaction of the cloud opacity with the gas opacity can mimic emission features in a non-scattering atmosphere. A non-scattering retrieval forward model will interpret it as such, resulting in an inverted thermal profile.  Despite the incorrectly retrieved TP profile, the non scattering cloud can accurately retrieve the abundances with a largely negligible bias.  

We find for the more realistic case of an atmosphere with a decreasing with height TP profile and a scattering cloud, a non-scattering cloud retrieval model cannot be distinguished between the scattering model. We find that for each wavelength range that retrieved temperature gradient for the non-scattering cloud model is smaller than the truth, with the opposite being true for the cloud free model. The retrieved abundances for the non-scattering cloud model are consistent with the scattering model for the individual instruments, however when combined for the "full" JWST range the abundances become biased by roughly 1 dex for both CO and H$_2$O in the non-scattering cloud case, and by roughly 1 dex for CO in the cloud free case.

\section{Non-constant cloud single-scattering albedo}
\label{Section5}

In the above proof of concept we have used a relatively simple setup, where we consider the single-scattering albedo and the opacity to be constant with wavelength. In this section we present some atmospheric simulations where we have used Mie theory to calculate the single-scattering albedo for some common cloud condensing species.  We present the single-scattering albedo as a function of particle size and wavelength for these common cloud species in Figure \ref{fig:real_ssa}, with  their properties taken from \citet{Kitzmann2018} apart from S$_8$, which was taken from Fuller, K. A. et al. in \citet{PalikEdwardD1998Hooc}. 

We can see from Figure \ref{fig:real_ssa} that the single-scattering albedo of the majority of the molecules either follow a straight line or a step like shape. This motivated our three-point cloud parameterisation scheme, as described in Section \ref{cloud param}. 

\subsection{Realistic cloud single-scattering albedo and constant opacity}
\label{real_const}

We explore whether our three-point cloud parameterisation scheme is able to account for the scattering properties of potential clouds without introducing bias in the retrieved abundances or thermal structure.  This three-point cloud approach is independent of the choice of specific condensate, making it a favorable approach for a generalized cloudy retrieval. We consider four cloud condensing species: Al$_2$O$_3$, MnS, MgSiO$_3$ and Fe, chosen as these condensates span the range of relevant optical morphologies. For each of the forward models we consider the particle radius to be of 10 $\mu$m and to have a variance of 0.1 following a Gamma distribution as shown in \citet{hansen1974light}. We consider the condensates to be vertically mixed, with a fractional scale height of 1 and to have abundance of log($\kappa_{\text{cld}}$) = 4. As in Section \ref{iso const}, we assume an isothermal atmosphere, resulting in "emission-like" features due to the scattering itself. Simulated observations are produced using the same procedure described in Section \ref{Section4} for the "full" wavelength range.

We then retrieve on the four different condensate set ups with three different models. We use the analytical pressure-temperature profile from Section \ref{TP_param} and an atmosphere with H$_2$O and CO gaseous absorption with abundances the same as those used in Section \ref{Section4}. Additionally, each of our three retrievals treats the clouds with an increasing complexity:
\begin{enumerate}[1.]
    \item A cloud-free model, shown in blue in upcoming figures.
    \item A pure absorbing cloud model, shown in red in upcoming figures. 
    \item A three-point cloud model. This is our novel cloud parameterisation, which is discussed in Section \ref{cloud param}. This model is shown in purple in upcoming figures.
\end{enumerate}

Our results for the realistic SSA are similar to the ones with the constant SSA described in the previous section. All retrievals that do not take thermal scattering into account estimate biased abundances and converge towards the wrong thermal structure (see Fig. \ref{fig:al2o3_spec}, \ref{fig:mns_spec}, \ref{fig:mgsio3_spec} and \ref{fig:fe_spec}.)

\begin{figure*}
    \centering
    \includegraphics[width=0.99\textwidth]{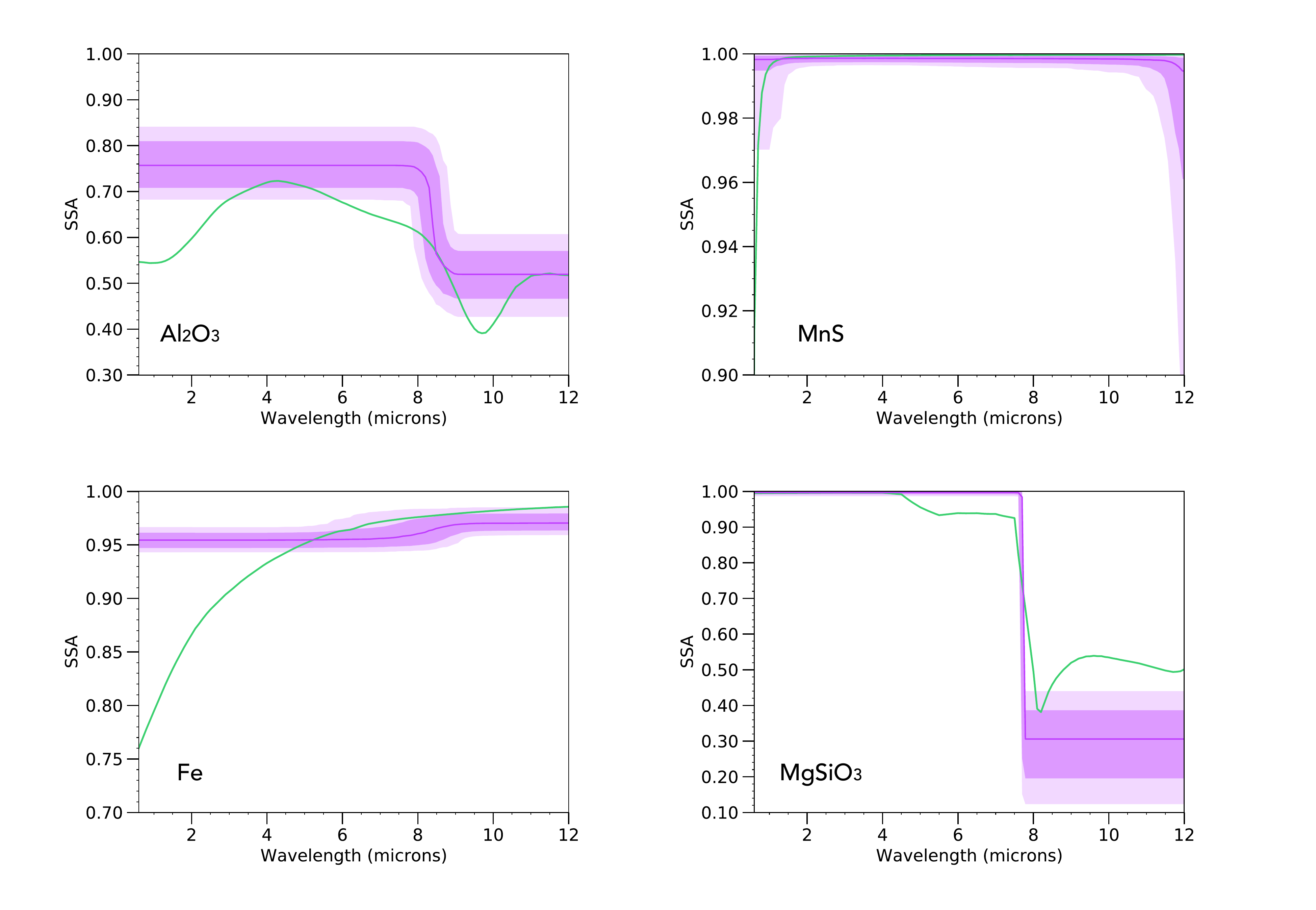}
    \caption{The best-fitting retrieved single-scattering albedo (SSA) spectrum using the three-point cloud parameterisation. In purple we present the retrieved SSA spectra with 1- and 2-sigma envelopes. In green we present the SSA spectrum for each of the cloud condensing species, which are labelled in each of the plots; each species has been modelled with a particle size of 10 $\mu$m.}
    \label{fig:ssa_grid}
\end{figure*}

Importantly, as shown in Fig. \ref{fig:ssa_grid}, we are able to retrieve the overall wavelength dependent shape of the the single-scattering albedo for all species considered. This is a very powerful result, given the approach was completely agnostic to the chemical and physical properties of the cloud, hence verifying that it works with a range of different shaped SSA spectra means that it is possible to learn about the clouds physical properties, this can be done by comparing the retrieved SSA spectrum with various different cloud condensing species.

\subsection{Realistic cloud single-scattering albedo spectrum and varying opacity}
We now investigate how the three-point cloud model would perform when retrieving on a spectrum that has a more physically motivated cloud. For this we use the \textsc{CHIMERA} \citep{line2013systematic} retrieval tool (\url{https://github.com/mrline/CHIMERA}) which employs the \citet{2001Ackerman} cloud model to produce realistic vertical droplet profiles over a broad range of droplet sizes (for a specific condensate), given a sedimentation efficiency (f$_{\text{sed}}$), eddy diffusivity (K$_{zz}$, simplistically assumed to be $10^{8}$ cm$^2$/s), cloud base pressure, and the condensate mixing ratio at the cloud base (see \citet{Mai2019} for implementation details). Here we assume enstatite (MgSiO$_3$) physical/optical properties, given a condensate mixing ratio of $10^{-5.5}$ at the cloud base (200 mbar).  Along with these cloud properties, we include gaseous opacity due to H$_2$O (4$\times$10$^{-4}$), CO (2$\times$10$^{-4}$), and H$_2$-H$_2$/He (0.86/0.137) CIA, and employ the analytical TP profile as discussed in Section \ref{TP_param} (T$_{\text{irr}}$ = 1400, log$\kappa_{\text{ir}}$ = -1.5, log$\gamma_1$ = -0.7, T$_{\text{int}}$ = 300K), under WASP-43b system properties.  Molecular (and continuum/condensate) opacities are treated within the correlated-K resort-rebin framework (e.g., \citet{Molliere2015,Amundsen2017}).  The \citet{Toon1989} two stream source function technique is used to solve for the fluxes (both up, down, stellar, and planetary) in these non-homogeneous multiple scattering atmospheres (a different approach to the multiple scattering problem than \textsc{NEMESIS}).  While stellar reflected flux is a natural consequence of multiple-scattering atmospheres, here, we explore only the retrieval biases arising from the thermal component.  We assume an extended cloud scenario for these tests (f$_{\text{sed}}$ = 0.1, a nearly constant with altitude condensate profile).

We then used this forward model in \textsc{PandExo} to generate the observations for one eclipse observation of WASP-43b using NIRSpec Prism and MIRI LRS configurations at a resolution of R = 50. These simulations are similar to the ones calculated in Section \ref{Section4}.  However, for this scenario we have kept the random noise instance to emulate a "real observation". We then retrieve on these data with the same three models described in Section \ref{real_const}. 

To validate \textsc{CHIMERA} and \textsc{NEMESIS} for this work we performed a benchmark test. We used \textsc{NEMESIS} to retrieve on a cloud free model generated by \textsc{CHIMERA} with an error envelope of 60ppm. We find consistent results as shown in Fig. \ref{fig:chivnem_spec} and \ref{fig:chivnem_triangle}. Therefore the differences that are found due to the 3 parameter SSA modelling approach and the Ackerman $\&$ Marley cloud parameterisation are not due to the choice of radiative transfer assumptions in the respective codes.

\subsubsection{Retrieval results}
\begin{figure*}
    \centering
    \includegraphics[width=\textwidth]{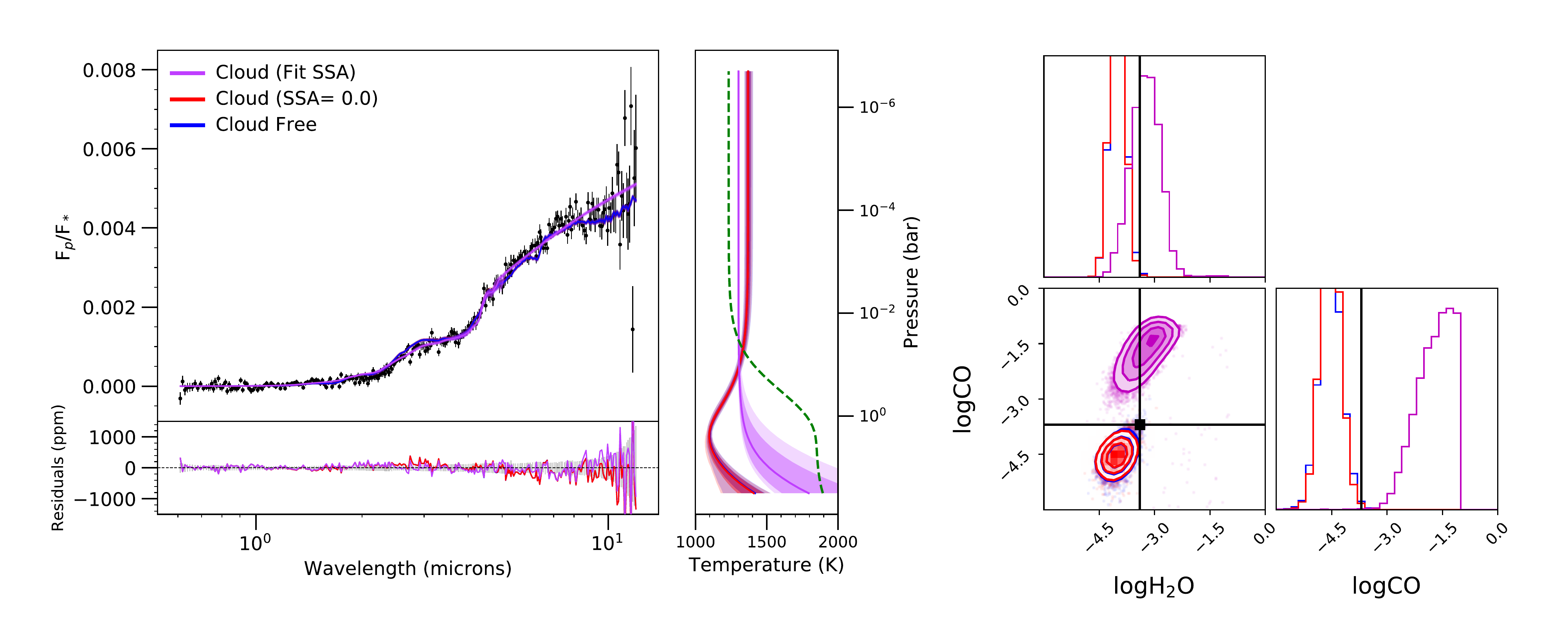}
    \caption{The retrieved results from the data generated using \textsc{CHIMERA}, the error bars and noise were generated using \textsc{PandExo}. Left: The best fitting spectra and retrieved temperature profiles. In purple we present the 3-point cloud model. In red we present the purely absorbing cloud model. In blue we present the cloud free model. In green we present the input temperature profile. Right: The retrieved chemical abundances, with the true value represented by a black line.}
    \label{fig:mike_pandexo}
\end{figure*}

Figure \ref{fig:mike_pandexo} shows the best fitting models, their temperature profiles and abundances. The purely absorbing cloud and cloud free retrievals produce nearly identical fits to the simulated data. This is due to the pure absorbing cloud model fitting for a near zero opacity which is consistent with a cloud-free atmosphere. Both models fit for an inverted TP profile and retrieve abundances that are lower than the input model.

The three point cloud model is able to retrieve the correct H$_2$O abundance, however the CO abundance is nearly 2 order of magnitudes larger than the input. This model is able to retrieve a temperature profile that is decreasing with height. Despite the CO abundance, the retrieved results are more consistent with the input, therefore, the scattering cloud model is an improvement compared to the non scattering and cloud free models. We find that the retrieve SSA spectrum is representative of the SSA spectra of the photosphere of the forward modelled atmosphere which is around 10$^{-2}$ to 10$^{-3}$ (Fig. \ref{fig:ssa_chimera}). However, the change of SSA with altitude, related to a change of particle size with altitude cannot be taken into account in our modelling framework and is probably leading to the bias in the retrieved abundances. 

We compute the Bayesian evidence to conclude which model is most supported by the data. We find that the data supports the 3-point scattering model over the purely absorbing and cloud free models by 8.25-$\sigma$ and 8.00-$\sigma$ respectfully. 

We note that the model used here to test our framework is probably a bit pessimistic. The nightside of the real WASP-43b planet is likely to have a larger thermal gradient than predicted by the semi-grey thermal profile, which should lead to larger spectral features than the ones predicted here.  

We note that our 3 parameter SSA method offers an improvement over cloud free or purely absorbing clouds, but does not offer a perfect solution.  This is likely due to the additional simplifications in our cloud model, such as the constant-with-altitude mixing and different choice of particle size distribution.  Future work will focus on implementing more realistic vertical cloud droplet profiles. A hybrid approach between an Ackerman $\&$ Marley type vertical droplet parameterisation and an "N-parameter" SSA retrieval may provide a better representation of "realistic" clouds.  We also do not consider the influence of inhomogeneities (e.g., "holes" or "patchy clouds") which could further complicate thermal emission inferences. 

\section{Conclusion}

Clouds have been found to be common in exoplanet atmospheres and with upcoming advanced space based facilities such as JWST we need to understand how best to account for them in our retrieval models. In this study we have explored the impact that a scattering cloud has on the interpretation of an emission spectrum as observed with JWST. We summarise our conclusions as follows:
\begin{enumerate}[1.]
    \item The presence of a scattering cloud reduces the emissivity of the atmosphere and leads to brightness temperatures that are smaller than real temperatures, this can been seen from Eq \ref{t_bright}. We further show that even when the cloud has a constant single-scattering albedo, the emissivity of the atmosphere varies with wavelength. This results in spectral features being present due to the cloud even if the atmosphere is isothermal. Finally we show that the variation of SSA with wavelength can introduce additional cloud features in the emission spectrum.
    \item We show that using a pure absorbing cloud, which is commonly used when interpreting clouds in exoplanet atmospheres, would lead to the inference of a systematically smaller than truth temperature gradient in an atmosphere with a non-inverted temperature profile. In the specific case of an isothermal atmosphere, the emission features produced by the thermal scattering would lead to the spurious detection of a thermal inversion. Similarly we find that when the scattering is not taken into account, we retrieve the incorrect chemical abundance for the molecules in the atmosphere. This is important, as the fits from the pure absorbing cloud and the clouds that account for the scattering are of similar quality. Hence, if the scattering was not considered, we would be tricked into believing that the exoplanet atmosphere has a thermal inversion and the accompanying incorrect chemical abundances.
    \item We develop a new three-point cloud parameterisation that can be used as an agnostic approach to fitting for clouds in emission. We show that the chemical abundance for a range of atmospheres with different cloud condensing species can be retrieved without assuming anything about the cloud. This is an improvement on what current models do, where an assumption about the cloud species is made before performing the retrieval \citep{Venot2020}, this assumption could bias the results. 
    \item We show that the three-point cloud parameterisation can also retrieve the general shape of the single-scattering albedo spectrum of the cloud. Despite the approach being agnostic to what sort of cloud is present, the retrieved single-scattering albedo can be compared to the single-scattering albedo of real species to infer the range of possible chemical composition and particle size distribution of the cloud.
    \item We find that our simple 3-point model cannot completely capture the complexities of the Ackerman $\&$ Marley model. This is due to our constant-with-altitude particle size and mixing. Future improvements on this work will include a hybrid approach between the "N-parameter" SSA retrieval and the vertical droplet parameterisation of the Ackerman $\&$ Marley model. Additionally, the spectrum used to test our framework is possibly pessimistic, as the real planet will probably have a larger thermal gradient and thus show larger spectral features.
\end{enumerate}

\section*{Acknowledgements}

Jake Taylor is a Penrose Graduate Scholar and would like to thank the Oxford Physics Endowment for Graduates (OXPEG) for funding this research. We thank Yakari for entertaining Vivien Parmentier's children while working on this manuscript from home. The majority of this manuscript was written while in lockdown due to COVID-19 and it would not have been possible to complete this study without the encouragement and support that Jake received from his supervisors Vivien Parmentier, Patrick Irwin and Suzanne Aigrain. Jake would also like to thank everyone at the University of Oxford's sub department of Atmospheric, Oceanic and Planetary Physics for providing loads of online support and being there for each other during this difficult time. We would like to thank the anonymous reviewer, their comments greatly improved our manuscript.

\section*{Data Availability}
The models generated in this paper are available on request.
%%%%%%%%%%%%%%%%%%%%%%%%%%%%%%%%%%%%%%%%%%%%%%%%%%

%%%%%%%%%%%%%%%%%%%% REFERENCES %%%%%%%%%%%%%%%%%%

% The best way to enter references is to use BibTeX:

\bibliographystyle{mnras}
\bibliography{bibliography} % if your bibtex file is called example.bib

%%%%%%%%%%%%%%%%%%%%%%%%%%%%%%%%%%%%%%%%%%%%%%%%%%

%%%%%%%%%%%%%%%%% APPENDICES %%%%%%%%%%%%%%%%%%%%%

\appendix
\section{An Analytical Interpretation}
\label{general_rutten}
In this section we attempt to summarise the general case. For a more in depth look, we recommend the work in \citet{Rutten2003}, which greatly helped our understanding of how scattering affects the planetary flux.

\subsection{General case}
Let us first consider the radiative transfer equation (RTE) in plane-parallel geometry:
\begin{equation}
\label{RTE}
    \mu\frac{dI_{\lambda}}{d\tau_{\lambda}} = I_{\lambda} - S_{\lambda}
\end{equation}
here $d\tau_{\lambda} = -\kappa_{\lambda}\rho dr$ , where $\kappa_{\lambda}$ is the wavelength-dependent opacity (also known as the cross section per unit mass, with units cm$^2$g$^{-1}$) and $\rho$ is the density in units cm$^{-3}$g. The intensity is $I_{\lambda}$ and the source function is $S_{\lambda}$.

If we consider that the source function is isotropic and average over all angles we get:

\begin{equation}
\label{angle_average_1}
    \frac{1}{2} \int_{-1}^{+1}\mu\frac{dI_{\lambda}}{d\tau_{\lambda}}d\mu = \frac{1}{2} \int_{-1}^{+1}I_{\lambda}d\mu - \frac{1}{2} \int_{-1}^{+1}S_{\lambda}d\mu
\end{equation}
By considering the different moments of intensity we respect to $\mu$ we can write this equation as:

\begin{equation}
\label{transport1}
    \frac{dH_{\lambda}}{d\tau_{\lambda}} = J_{\lambda} - S_{\lambda}
\end{equation}

Where $H_{\lambda}$ is the second order moment, otherwise know as the Eddington flux; it can be related to the flux by $H_{\lambda} = \frac{F_{\lambda}}{4}$. $J_{\lambda}$ is the first order moment, otherwise know as the mean intensity. 

If we multiply Equation \ref{RTE} by $\mu$ and then average over all angles we instead get:

\begin{equation}
    \frac{1}{2} \int_{-1}^{+1}\mu^{2}\frac{dI_{\lambda}}{d\tau_{\lambda}}d\mu = \frac{1}{2} \int_{-1}^{+1}\mu I_{\lambda}d\mu - \frac{1}{2} \int_{-1}^{+1}\mu S_{\lambda}d\mu
\end{equation}

which can be written in terms of the moments on intensity as:

\begin{equation}
\label{transport2}
    \frac{dK_{\lambda}}{d\tau_{\lambda}} = H_{\lambda}
\end{equation}

Here $K_{\lambda}$ is the third-order moment of the intensity and is known as the $K$ integral. We lose the source term as this goes to zero due to it being isotropic. We can then use Equation \ref{transport1} and Equation \ref{transport2} to get:
\begin{equation}
\label{SORTE}
    \frac{d^{2}K_{\lambda}}{d\tau_{\lambda}^{2}} = J_{\lambda} - S_{\lambda}
\end{equation}
Hence, we now have a second order differential version of the radiative transfer equation. 

We can now relate $K_{\lambda}$ with $J_{\lambda}$ by considering the following:
\begin{equation}
   K_{\lambda} = \frac{1}{2} \int_{-1}^{+1}\mu^{2}I_{\lambda}d\mu
   = \frac{1}{2}<I_{\lambda}>\int_{-1}^{+1}\mu^{2}d\mu = \frac{1}{3}J_{\lambda}
\end{equation}
This relation is known as the first order Eddington approximation, using this we can write Equation \ref{SORTE} as:
\begin{equation}
    \frac{1}{3}\frac{d^{2}J_{\lambda}}{d\tau_{\lambda}^{2}} = J_{\lambda} - S_{\lambda}.
\end{equation}

If we consider that the scattering is elastic, the source function becomes
\begin{equation}
    S_{\lambda} = (1-\epsilon_{\lambda})J_{\lambda} - \epsilon_{\lambda}B_{\lambda}
\end{equation}
and we can then write the radiative transfer equation as
\begin{equation}
\label{K_relation}
    \frac{1}{3}\frac{d^{2}J_{\lambda}}{d\tau_{\lambda}^{2}} = \epsilon_{\lambda}(J_{\lambda} - B_{\lambda}).
\end{equation}

Let us now assume that the Planck function varies linearly with optical depth
\begin{equation}
    B_{\lambda}(\tau_{\lambda}) = B_{\lambda,0} + b_{\lambda}\tau_{\lambda}
\end{equation}
which results in the second order differential of the Planck function to be zero ($\frac{d^{2}B_{\lambda}}{d\tau_{\lambda}^{2}} = 0$) and therefore we can rewrite Equation \ref{K_relation} as
\begin{equation}
    \frac{1}{3}\Bigg(\frac{d^{2}J_{\lambda}}{d\tau_{\lambda}^{2}} - \frac{d^{2}B_{\lambda}}{d\tau_{\lambda}^{2}}\Bigg) = \epsilon_{\lambda}(J_{\lambda} - B_{\lambda}).
\end{equation}
as this does not change the result of Equation \ref{K_relation}. We can simplify this by writing it in the form
\begin{equation}
    \frac{1}{3}\frac{d^{2}}{d\tau_{\lambda}^{2}}(J_{\lambda} - B_{\lambda}) = \epsilon_{\lambda}(J_{\lambda} - B_{\lambda})
\end{equation}
which can be integrated to find the difference between $J_{\lambda}$ and $B_{\lambda}$
\begin{equation}
\label{j_b_difference}
    J_{\lambda} - B_{\lambda} = C_1e^{-\sqrt{3\epsilon_{\lambda}}\tau_{\lambda}} + C_2e^{\sqrt{3\epsilon_{\lambda}}\tau_{\lambda}}
\end{equation}

where $C_1$ and $C_2$ are integration constants that can be found by applying a few boundary conditions. These boundary conditions are that $J_{\lambda} = B_{\lambda}$ for $\tau_{\lambda} \rightarrow \infty$ and no incident flux at $\tau_{\lambda} = 0$ hence in the Rosseland approximation. This results in $C_2 = 0$, to find $C_1$ we adopted the second Eddington approximation $J_{\lambda}(0) = a_{\lambda}H_{\lambda}(0)$, note that the second Eddington approximation is $J_{\lambda}(0) = 2H_{\lambda}(0)$ for an unrealistic Lambert radiator, we therefore use $a_{\lambda}$ as a free parameter.
\begin{equation}
\label{c_part1}
    J_{\lambda}(0) = B_{\lambda,0} + C_1
\end{equation}

\begin{equation}
\label{c_part2}
\begin{aligned}
    J_{\lambda}(0) = a_{\lambda}H_{\lambda}(0) = a_{\lambda}\Big[\frac{dK_{\lambda}}{d\tau_{\lambda}}\Big]_{\tau_{\lambda} = 0} = (a_{\lambda}/3)\Big[\frac{dJ_{\lambda}}{d\tau_{\lambda}}\Big]_{\tau_{\lambda} = 0} \\
    = -(a_{\lambda}/3)\sqrt{3\epsilon_{\lambda}}C_1 + a_{\lambda}b_{\lambda}/3
\end{aligned}
\end{equation}
By combining Equation \ref{c_part1} and \ref{c_part2} we find $C_1$ to be
\begin{equation}
\label{c_equation}
    C_1 = -\frac{B_{\lambda,0} - a_{\lambda}b_{\lambda}/3}{1 + (a_{\lambda}/3)\sqrt{3\epsilon_{\lambda}}}
\end{equation}
and if we substitute Equation \ref{c_equation} into \ref{j_b_difference} we can find the general scattering solutions for $J_{\lambda}(\tau_{\lambda})$, $S_{\lambda}(\tau_{\lambda})$ and $H_{\lambda}(\tau_{\lambda})$. Hence
\begin{equation}
    J_{\lambda}(\tau_{\lambda}) = B_{\lambda,0} + b_{\lambda}\tau_{\lambda} - \frac{B_{\lambda,0} - a_{\lambda}b_{\lambda}/3}{1 + (a_{\lambda}/3)\sqrt{3\epsilon_{\lambda}}}e^{-\sqrt{3\epsilon_{\lambda}}\tau_{\lambda}}
\end{equation}

\begin{equation}
    S_{\lambda}(\tau_{\lambda}) = B_{\lambda,0} + b_{\lambda}\tau_{\lambda} - (1 - \epsilon_{\lambda})\frac{B_{\lambda,0} - a_{\lambda}b_{\lambda}/3}{1 + (a_{\lambda}/3)\sqrt{3\epsilon_{\lambda}}}e^{-\sqrt{3\epsilon_{\lambda}}\tau_{\lambda}}
\end{equation}
\begin{equation}
    H_{\lambda}(\tau_{\lambda}) = b_{\lambda}/3 + \sqrt{\epsilon_{\lambda}}\frac{B_{\lambda,0} - a_{\lambda}b_{\lambda}/3}{\sqrt{3} + a_{\lambda}\sqrt{\epsilon_{\lambda}}}e^{-\sqrt{3\epsilon_{\lambda}}\tau_{\lambda}}
\end{equation}
As in \citet{Rutten2003} we assume $a_\lambda=\sqrt{3}$ and assume that the planetary flux is given by $F_p=4\pi H$, we can substitute in the value for $H_{\lambda}(\tau_{\lambda})$ we just found to get

\begin{equation}
    F_p=4\pi\left(\frac{b_\lambda}{3}-\frac{\sqrt{\epsilon_\lambda}}{1+\sqrt{\epsilon_\lambda}}b_\lambda+\frac{\sqrt{\epsilon_\lambda}}{1+\sqrt{\epsilon_\lambda}}\frac{1}{\sqrt{3}}B_\lambda(0)\right)
\end{equation}

which can be re-written as: 
\begin{equation}
    F_p=4\pi\frac{2\sqrt{\epsilon_\lambda}}{1+\sqrt{\epsilon_\lambda}}\left(B_\lambda(0)+\frac{b_\lambda}{\sqrt{3\epsilon_\lambda}}\right)
\end{equation}
Or also:
\begin{equation}
\label{p_flux}
    F_p=\frac{2\pi}{\sqrt{3}}\frac{2\sqrt{\epsilon_\lambda}}{1+\sqrt{\epsilon_\lambda}}B_\lambda(\tau_\lambda=1/\sqrt{3\epsilon_\lambda})
\end{equation}
\section{Analytical look at how emissivity changes the spectrum}
\label{Appendix B}
Let us consider the case where the emissivity is wavelength dependant ($\epsilon_{\lambda}$). We now wonder how the spectrum is changing with wavelength. First, we will write the planetary flux as shown in Eq \ref{p_flux}:
\begin{equation}
    F_p=\frac{2\pi}{\sqrt{3}}\frac{2\sqrt{\epsilon_\lambda}}{1+\sqrt{\epsilon_\lambda}}B_\lambda(T(\tau_\lambda=1/\sqrt{3\epsilon_\lambda}))\text{.}
\end{equation}
In order to obtain the slope of the spectra, which determines its shape, we differentiate this with respect to wavelength:
\begin{equation}
    \frac{dF_p}{d\lambda} = \frac{2\pi}{\sqrt{3}} \frac{2df(\epsilon_\lambda)}{d\lambda}B_\lambda + \frac{2\pi}{\sqrt{3}} 2f(\epsilon_\lambda)\frac{dB_{\lambda}}{d\lambda} 
\end{equation}
here we have set $f(\epsilon_\lambda) = \frac{\sqrt{\epsilon_\lambda}}{1+\sqrt{\epsilon_\lambda}} $ for simplicity. We can now rewrite this equation as:
\begin{equation}
\label{B4}
    \frac{dF_p}{d\lambda} = \frac{2\pi}{\sqrt{3}} \frac{2df(\epsilon_\lambda)}{d\lambda}B_\lambda + \frac{2\pi}{\sqrt{3}} 2f(\epsilon_\lambda)\frac{dB_{\lambda}}{dT}\frac{dT}{dP}\frac{dP}{d\lambda} \text{.}
\end{equation}
The photospheric pressure $P_{\lambda}$ is the pressure where $\tau_{\lambda}=1$ and can be written as:
\begin{equation}
    P_\lambda = \frac{\mu g}{\sigma_{i,\lambda} f_i}\frac{1}{\sqrt{3\epsilon_{\lambda}}} \text{,}
\end{equation}
where $f_i$ is the mixing ratio/abundance of a molecule and $\sigma_{i,\lambda}$ is the absorption cross-section. As emissivity decreases, the photospheric pressure becomes larger. The photospheric pressure varies with wavelength both because the opacity varies and because the emissivity varies. We assume that only one species is radiatively active, so $f_i$ is wavelength independent, which gives:
\begin{equation}
    \frac{dP_\lambda}{d\lambda} = -\frac{\mu g}{\sigma_{i,\lambda} f_i}\frac{1}{\sqrt{3\epsilon_{\lambda}}}\frac{dln(\sigma_{i,\lambda})}{d\lambda} -\frac{\mu g}{\sigma_{i,\lambda} f_i}\frac{1}{\sqrt{3\epsilon_{\lambda}}}\frac{dln(\epsilon_{\lambda})}{d\lambda} \text{,}
\end{equation}
which can be simplified as:
\begin{equation}
\label{B7}
    \frac{dP_\lambda}{d\lambda} = -P_\lambda \frac{dln(\sigma_{i,\lambda} \epsilon_\lambda)}{d\lambda} \text{.}
\end{equation}
Next we need to find out how the function $f(\epsilon_\lambda) = \frac{\sqrt{\epsilon_\lambda}}{1+\sqrt{\epsilon_\lambda}} $ varies with wavelength:
\begin{equation}
\label{B8}
    \frac{df}{d\lambda} = \frac{1}{2(1+\sqrt{\epsilon_\lambda})^2\sqrt{\epsilon_\lambda}}\frac{d\epsilon_\lambda}{d\lambda} \text{.}
\end{equation}
Now we can insert Eq \ref{B7} and \ref{B8} into Eq \ref{B4} to determine how the emissivity impacts the planetary flux:
\begin{equation}
\begin{split}
    \frac{dF_p}{d\lambda} = \frac{2\pi}{3}\frac{1}{2(1+\sqrt{\epsilon_\lambda})^2\sqrt{\epsilon_\lambda}}B_\lambda(T_p)\frac{d\epsilon_\lambda}{d\lambda} + \\
    \frac{2\pi}{3}\frac{\sqrt{\epsilon_\lambda}}{1+\sqrt{\epsilon_\lambda}}\frac{dB}{dT}\frac{dT}{dP}\Big(-P_\lambda \frac{dln(\sigma_{i,\lambda} \epsilon_\lambda)}{d\lambda}\Big)
\end{split}
\end{equation}
which we can simplify to:
\begin{equation}
\begin{split}
\label{B9}
    \frac{dF_p}{d\lambda} = \frac{2\pi}{3}B_\lambda(T_p)\Big(\frac{1}{2(1+\sqrt{\epsilon_\lambda})^2\sqrt{\epsilon_\lambda}}\frac{d\epsilon_\lambda}{d\lambda} - \\
    \frac{\sqrt{\epsilon_\lambda}}{1+\sqrt{\epsilon_\lambda}}\frac{dlnB}{dlnT}\frac{dlnT}{dlnP}\frac{dln(\sigma_{i,\lambda} \epsilon_\lambda)}{d\lambda}\Big) \text{.}
\end{split}
\end{equation}

This equation contains two main terms. The first term is due to the fact that planetary flux at a given wavelength scales directly with the emissivity. So if emissivity increases with wavelength, the flux will also increase with wavelength. This term is independent of the temperature gradient and is thus the reason why spectral features can appear even when an isothermal thermal structure is used. For a planet with a constant scattering cloud, the emissivity would be written as:
\begin{equation}
\epsilon_{\lambda}=\frac{\sigma_{\rm gas}}{\sigma_{\rm gas}+\frac{f_{\rm clouds}}{f_{\rm gas}}\sigma_{\rm cloud}}
\end{equation}
which is a monotonic function of $\sigma_{\rm gas}$. As a consequence, this first term will produce emission features from the gaseous components when a scattering cloud is present in the atmosphere, whatever the thermal profile.

The second term is more complicated. When there is no scattering present it simply states that a non-inverted profile will produce absorption features and an inverted profile will produce emission features. Given that the variations of $\sigma$ and $\epsilon$ follow each other, the presence of scattering in an atmosphere will reinforce this behavior. So the presence of a purely scattering cloud will naturally increase the size of the emission and absorption features present.

Based on our numerical experiments shown in Figure \ref{fig:compare_ssa} we can see that the first term is usually larger than the second term. As a consequence, for a non-inverted thermal profile the presence of a purely scattering cloud will tend to reduce the absorption features as the two terms compensate each other. For an inverted thermal profile, the two terms act together and a scattering clouds will always enhanced the emission features.

%It can be seen that for an atmosphere with no inversion ($\frac{dT}{dP} > 0$) a large $\epsilon_\lambda$ would result in a large flux, however the large $\epsilon_\lambda$ results in smaller pressure layers being probed which would have the opposite effect. These should act to cancel each other out. Experimentally we show that the larger $\epsilon_\lambda$ results in more flux, so this is the dominating effect. For an atmosphere with an inversion ($\frac{dT}{dP} < 0$) the effects are combined, where the smaller the $\frac{dT}{dP}$ the stronger the effect.
\section{Plots for various simulations}

\begin{figure*}
    \centering
    \includegraphics[width=\textwidth]{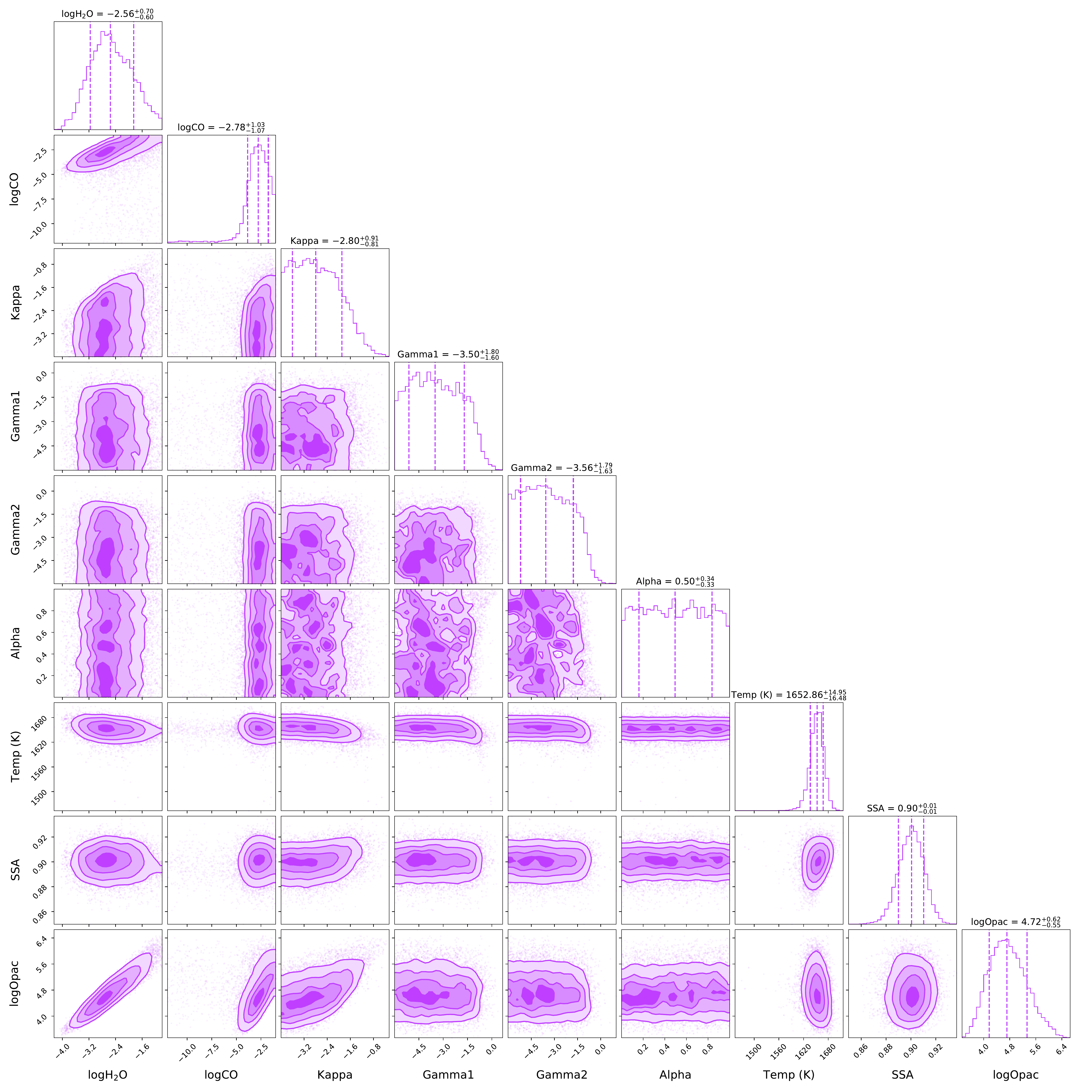}
    \caption{We present the corner plot showing the distribution of parameter space explored for the scattering retrieval presented in Fig. \ref{fig:ALL_Iso_Pandexo}. For the 1-D histograms, we present the median result and the 1-$\sigma$ values. This are represented by vertical dashed lines.   }
    \label{fig:iso_cloud_posterior}
\end{figure*}

\begin{figure*}
    \centering
    \includegraphics[width=\textwidth]{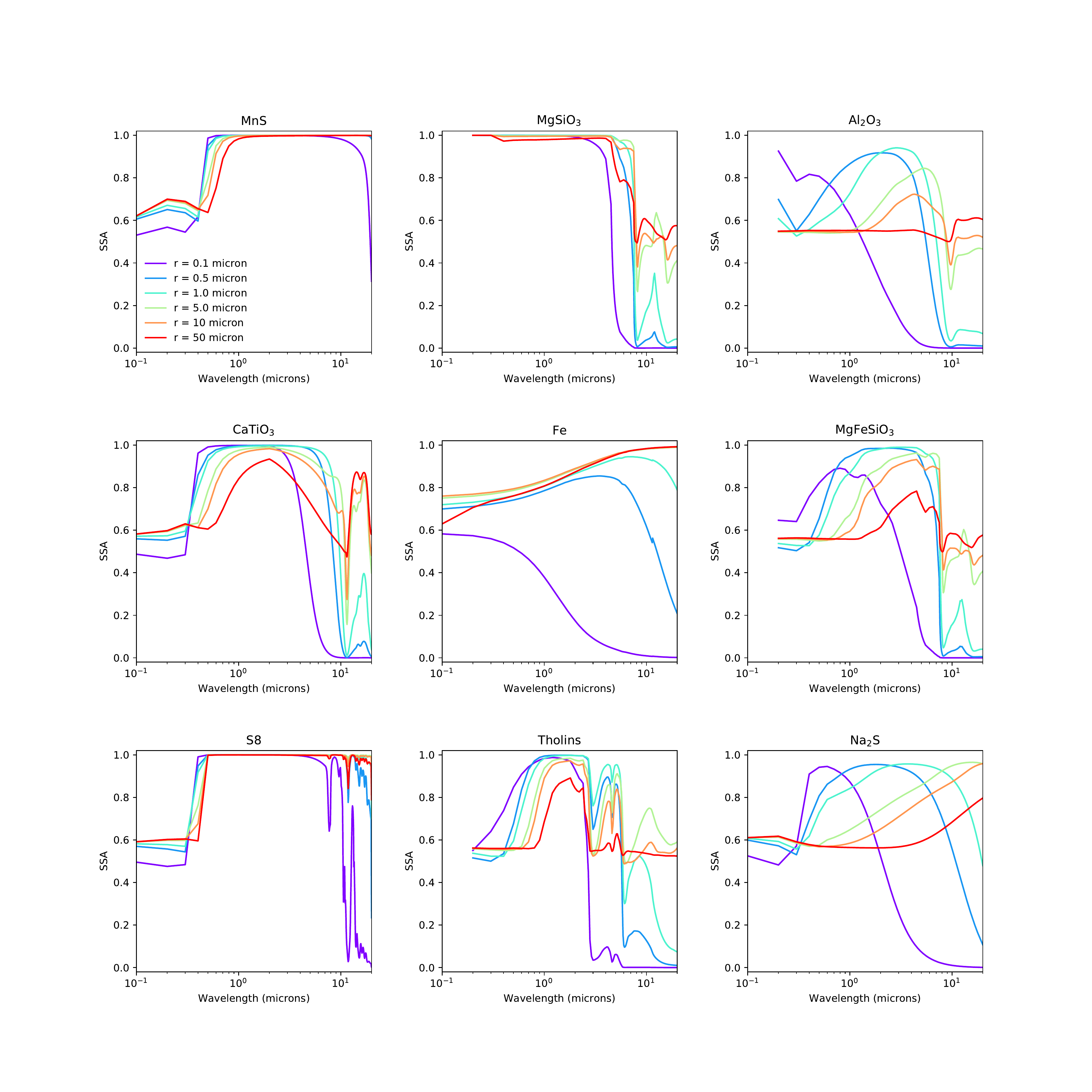}
    \caption{Common cloud species found in exoplanet atmospheres and their single-scattering albedos. They are calculated for particle radii of 0.1, 0.5, 1.0, 5.0, 10.0 and 50.0 $\mu$m over a wavelength range of 0.1 -- 20 $\mu$m for MnS, S$_8$, Fe and CaTiO$_3$ and a wavelength range of 0.2 - 20 $\mu$m for MgSiO$_3$, MgFeSiO$_3$ and Al$_2$O$_3$. Caution needs to be taken when interpreting the results from MnS as the optical constants used in the infrared were inferred from ZnS and Na$_2$S rather than laboratory experiments \citep{Kitzmann2018}.   }
    \label{fig:real_ssa}
\end{figure*}

\begin{figure*}
    \centering
    \includegraphics[width=\textwidth]{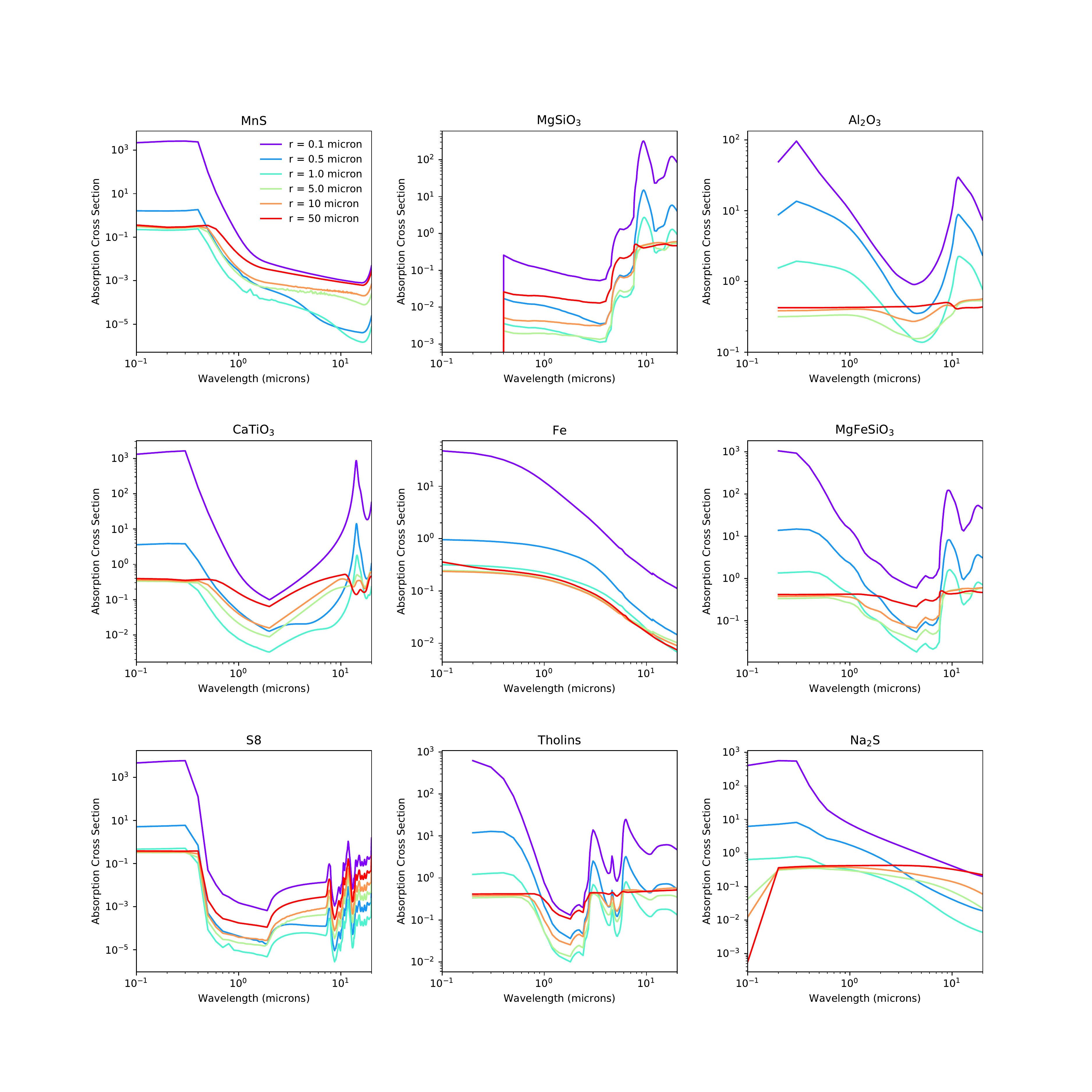}
    \caption{Common cloud species found in exoplanet atmospheres and their absorption cross sections. They are calculated for particle radii of 0.1, 0.5, 1.0, 5.0, 10.0 and 50.0 $\mu$m over a wavelength range of 0.1 - 20 $\mu$m for MnS, S8, Fe and CaTiO$_3$ and a wavelength range of 0.2 - 20 $\mu$m for MgSiO$_3$, MgFeSiO$_3$ and Al$_2$O$_3$. Caution needs to be taken when interpreting the results from MnS as the optical constants used in the infrared were inferred from Zns and Na2S rather than laboratory experiments \citep{Kitzmann2018}.  }
    \label{fig:real_abs}
\end{figure*}

\begin{figure*}
    \centering
    \includegraphics[width=\textwidth]{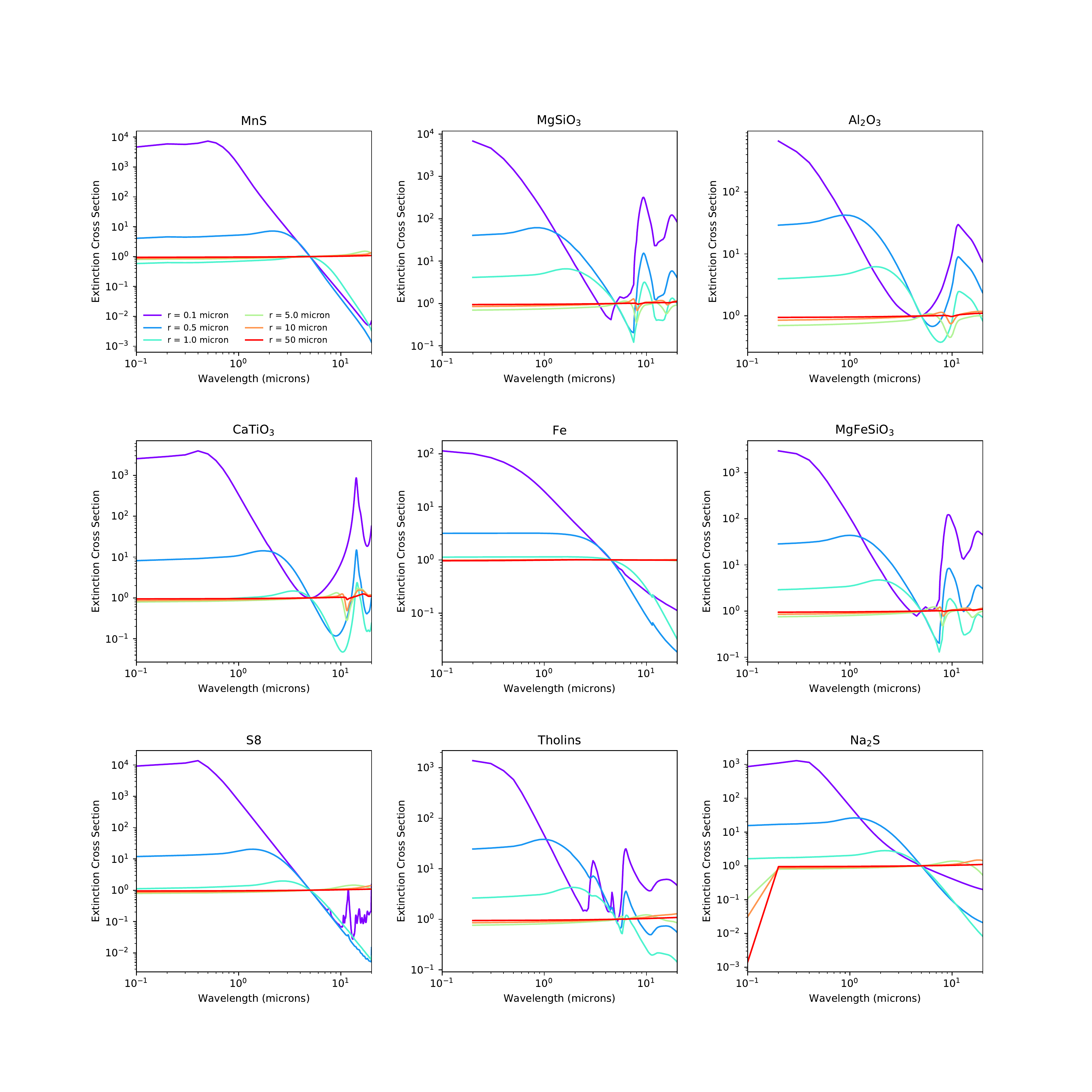}
    \caption{Common cloud species found in exoplanet atmospheres and their extinction cross section, which has been normalised to 1 at 5 $\mu$m, and are therefore unit less. They are calculated for particle radii of 0.1, 0.5, 1.0, 5.0, 10.0 and 50.0 $\mu$m over a wavelength range of 0.1 - 20 $\mu$m for MnS, S8, Fe and CaTiO$_3$ and a wavelength range of 0.2 - 20 $\mu$m for MgSiO$_3$, MgFeSiO$_3$ and Al$_2$O$_3$. Caution needs to be taken when interpreting the results from MnS as the optical constants used in the infrared were inferred from Zns and Na2S rather than laboratory experiments \citep{Kitzmann2018}.  }
    \label{fig:real_ext}
\end{figure*}

\begin{figure*}
    \centering
    \includegraphics[width=0.99\textwidth]{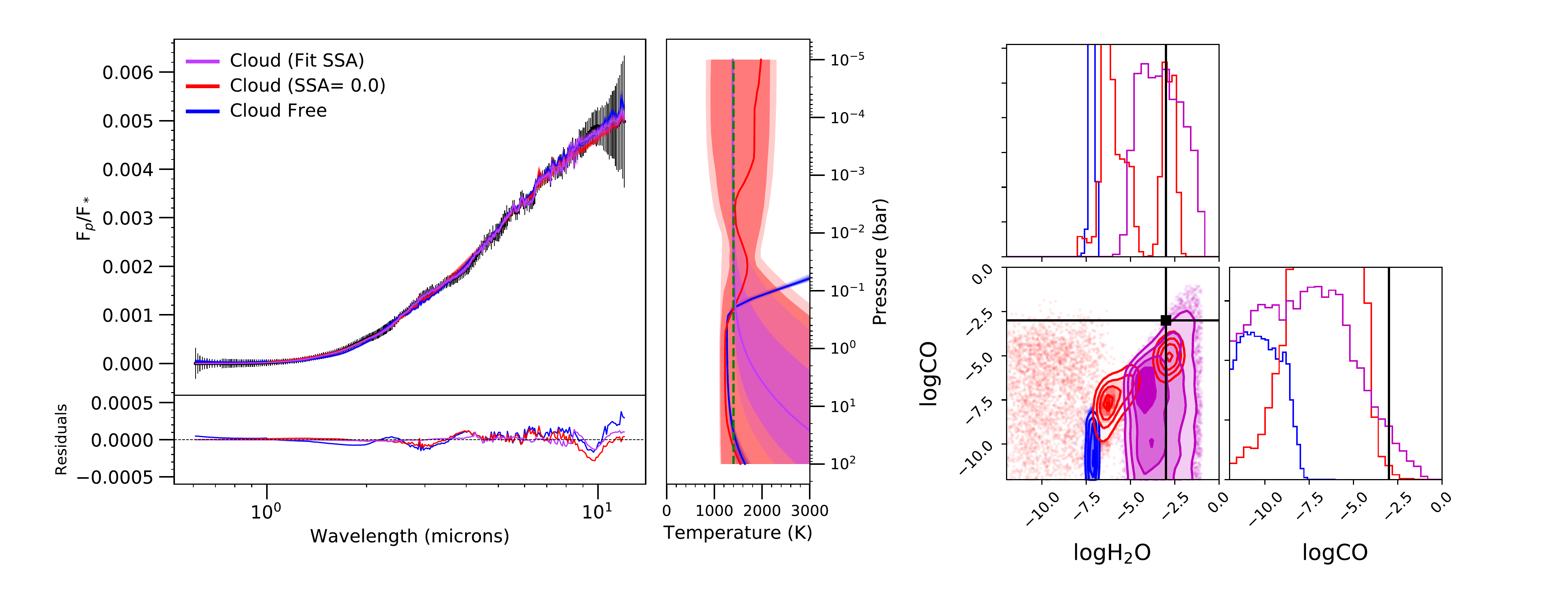}
    \caption{\textbf{Left:} The best-fitted spectra for the three models used on the data generated using a cloud made of Al$_2$O$_3$. For the cloud model we consider the particle radius to be of 10 $\mu$m and to have a variance of 0.1 following a Gamma distribution as shown in \citet{hansen1974light}. In purple we present the new three-point cloud, in red we present a cloud with single-scattering albedo of 0 (i.e., a grey cloud) and in blue we present the cloud-free model. Below we present the residuals of each fit. We present the accompanying temperature retrieval for each of the spectra, which have 1-$\sigma$ and 2-$\sigma$ shaded regions. The green dashed line is the input temperature-pressure profile. \textbf{Right:} We present the chemistry posterior for each case with the true value present in black. }
    \label{fig:al2o3_spec}
\end{figure*}

\begin{figure*}
    \centering
    \includegraphics[width=0.99\textwidth]{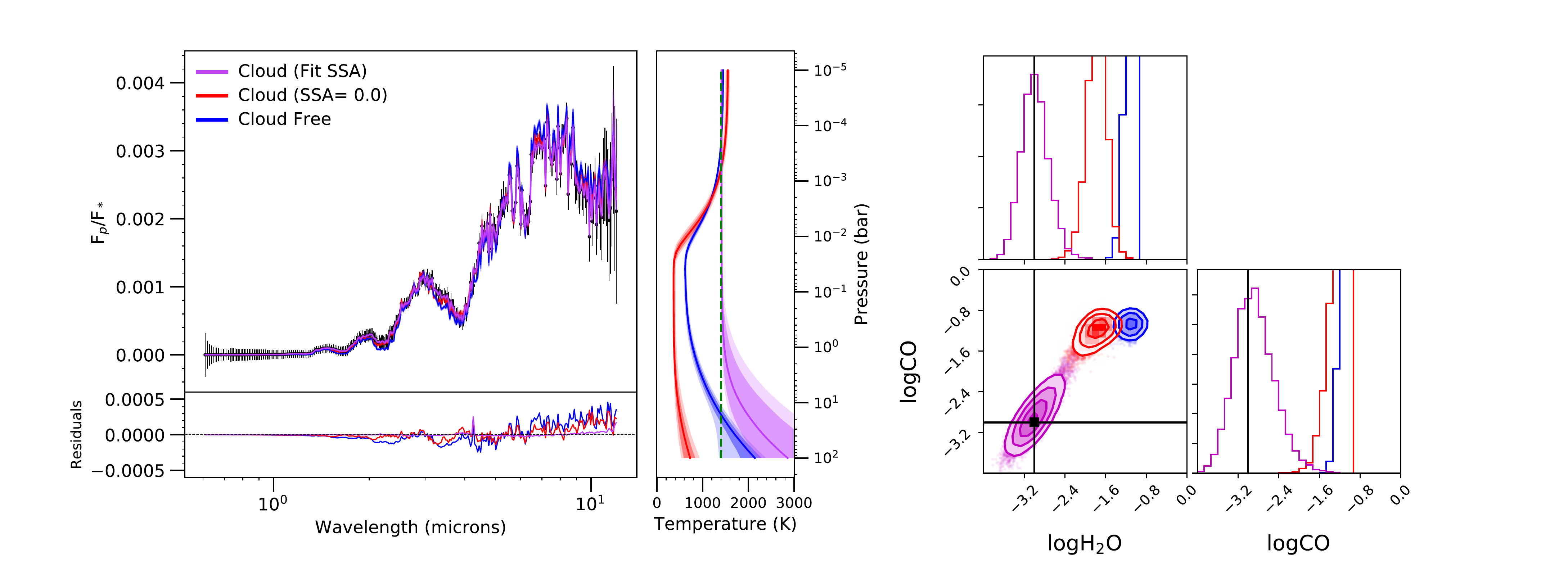}
    \caption{The same simulations as Figure \ref{fig:al2o3_spec}, but now for the case of an MnS cloud.}
    \label{fig:mns_spec}
\end{figure*}

\begin{figure*}
    \centering
    \includegraphics[width=0.99\textwidth]{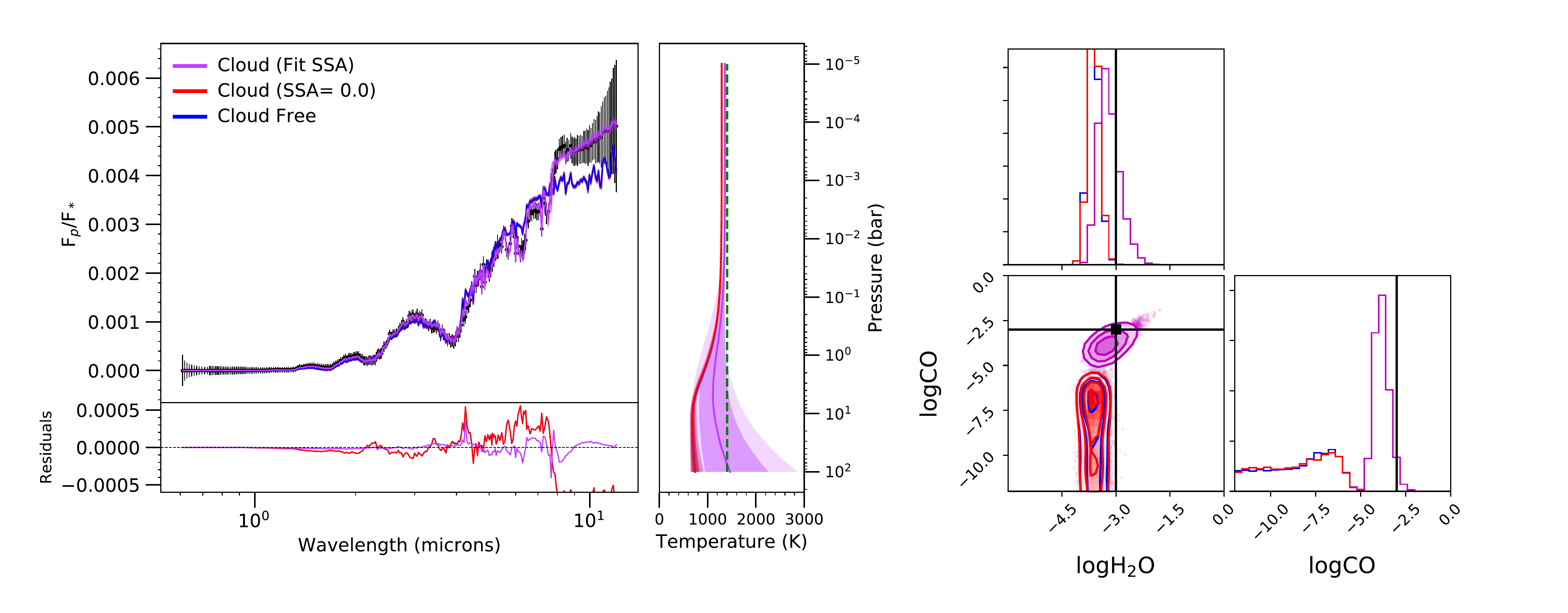}
    \caption{The same simulations as Figure \ref{fig:al2o3_spec}, but for the MgSiO$_3$ cloud case.}
    \label{fig:mgsio3_spec}
\end{figure*}

\begin{figure*}
    \centering
    \includegraphics[width=0.99\textwidth]{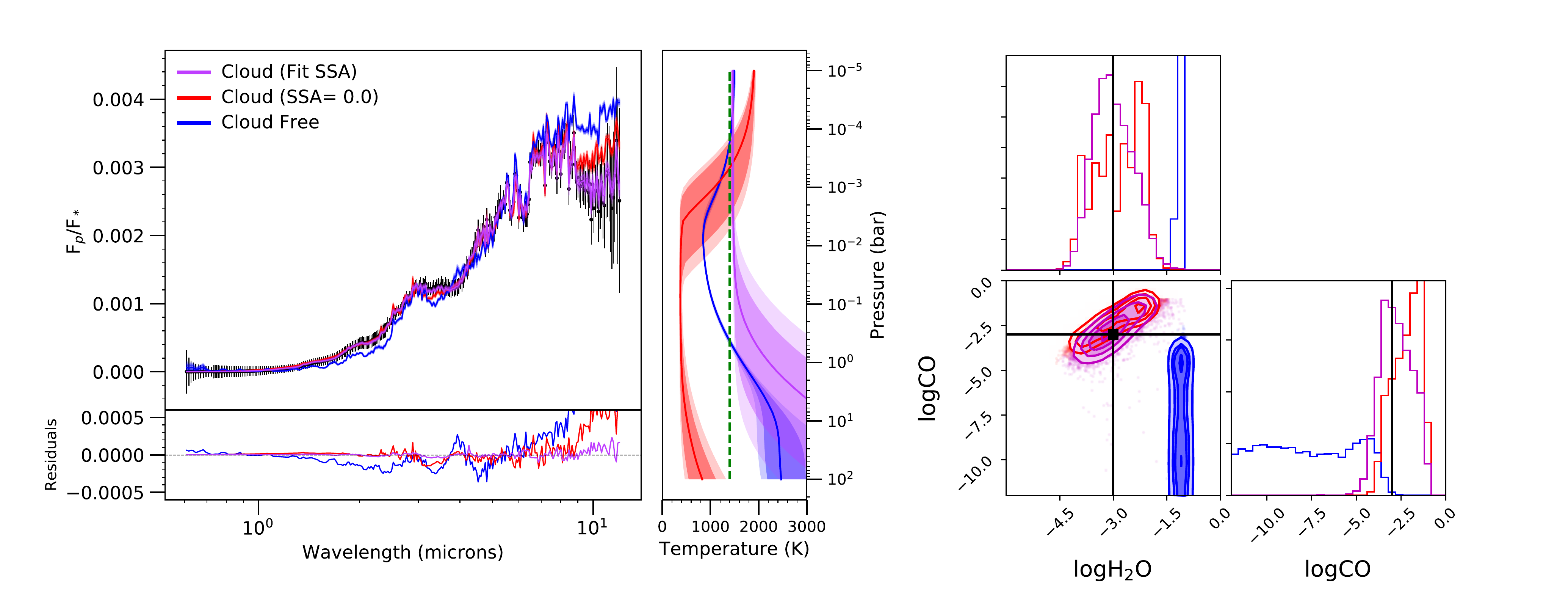}
    \caption{The same simulations as Figure \ref{fig:al2o3_spec},  but for the Fe cloud case.}
    \label{fig:fe_spec}
\end{figure*}

\begin{figure*}
    \centering
    \includegraphics[width=0.99\textwidth]{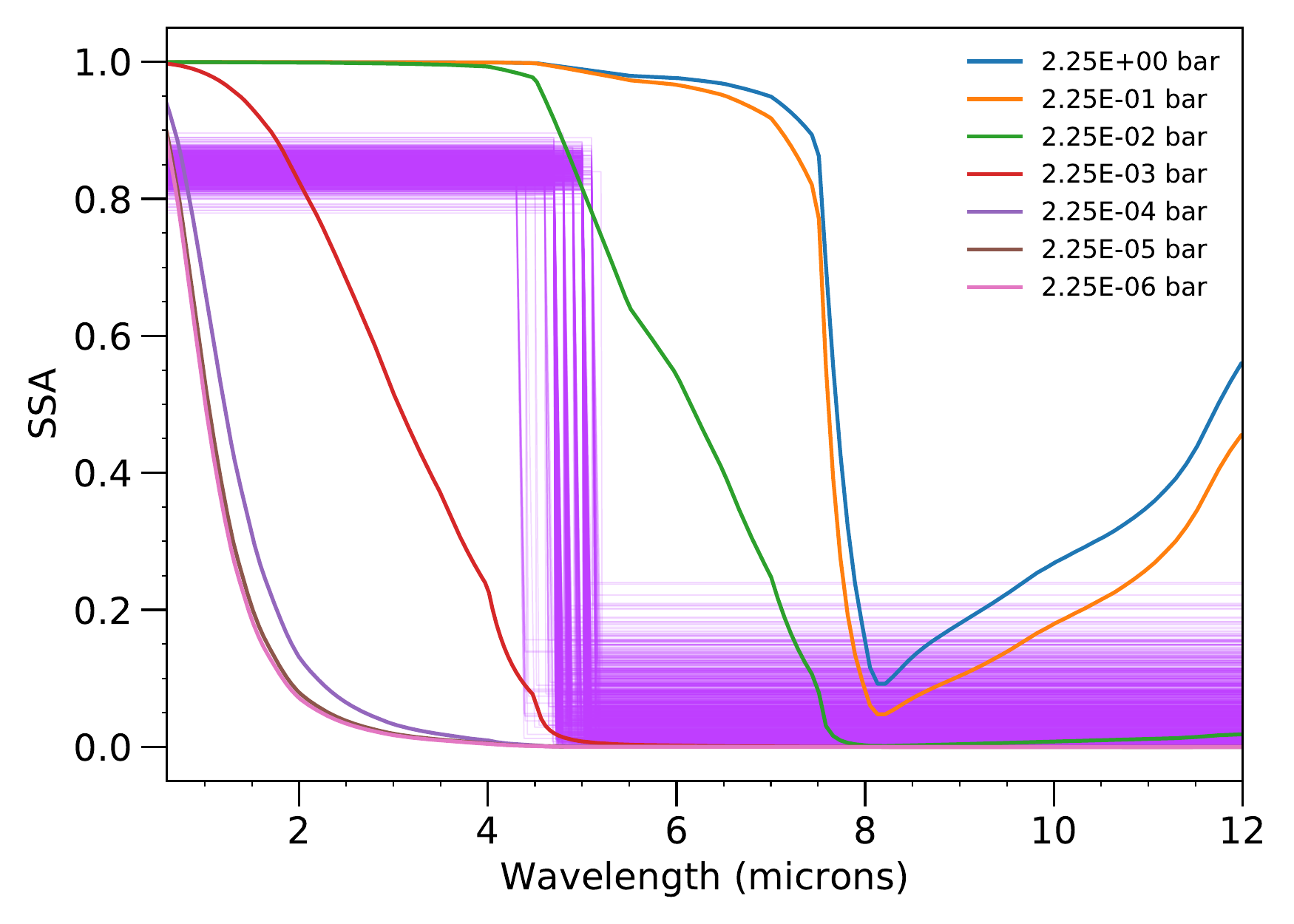}
    \caption{We present the retrieved single-scattering albedo for the 3-point model from Fig. \ref{fig:mike_pandexo}. The purple lines shows 1000 models drawn from the posterior distribution. We over plot the forward model single-scattering albedo generated using \textsc{CHIMERA} for different pressure levels. We note that the photosphere is between 10$^{-2}$ and 10$^{-3}$ bar, our retrieved single-scattering albedo spectrum is between the forward model single-scattering albedo spectra at these values. }
    \label{fig:ssa_chimera}
\end{figure*}

\begin{figure*}
    \centering
    \includegraphics[width=0.99\textwidth]{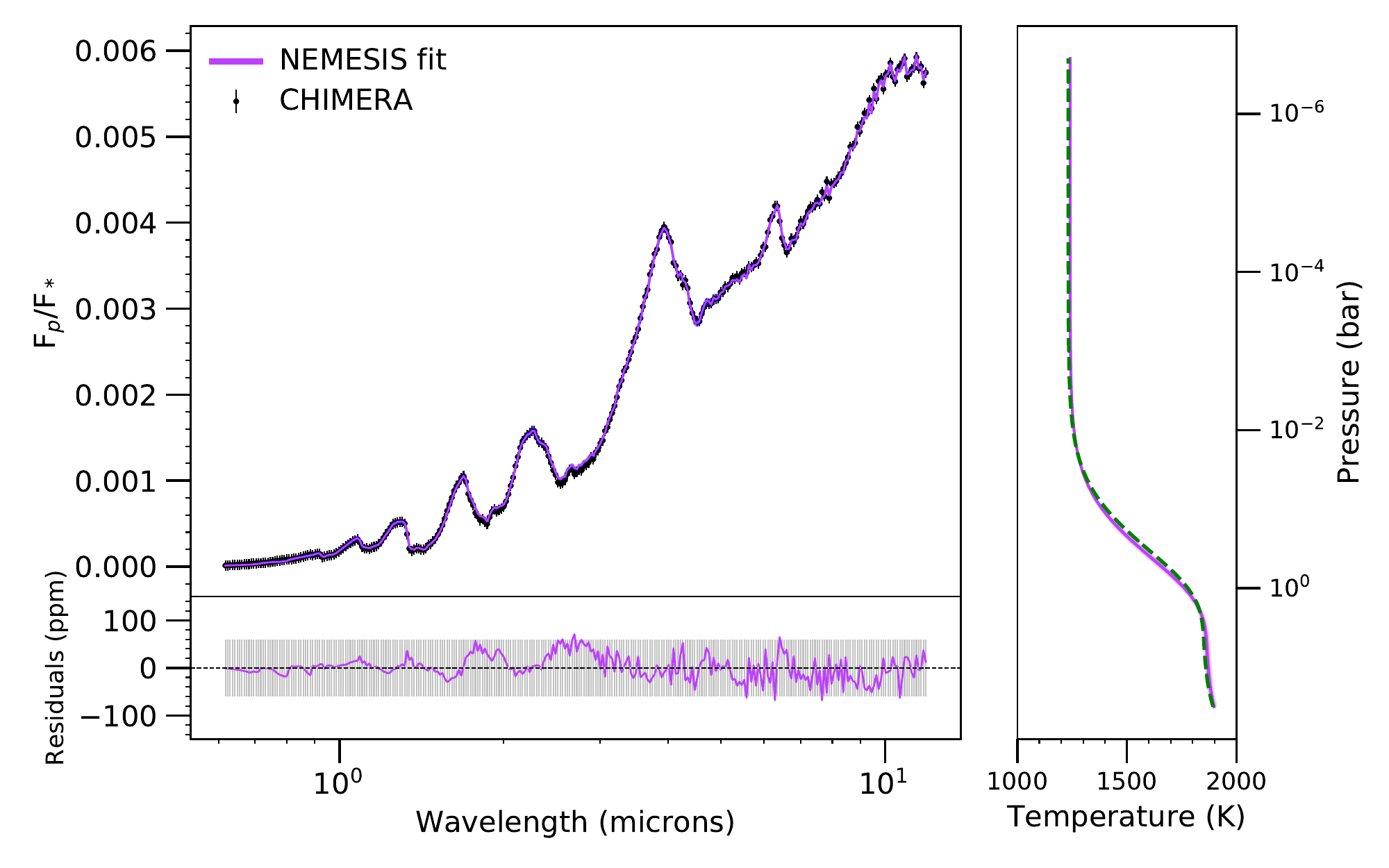}
    \caption{Retrieved results by performing a cloud free benchmark test on a model generated using \textsc{CHIMERA}. In purple we show the best fitting models from \textsc{NEMESIS}. We show the input TP structure by a green dashed line.}
    \label{fig:chivnem_spec}
\end{figure*}

\begin{figure*}
    \centering
    \includegraphics[width=0.99\textwidth]{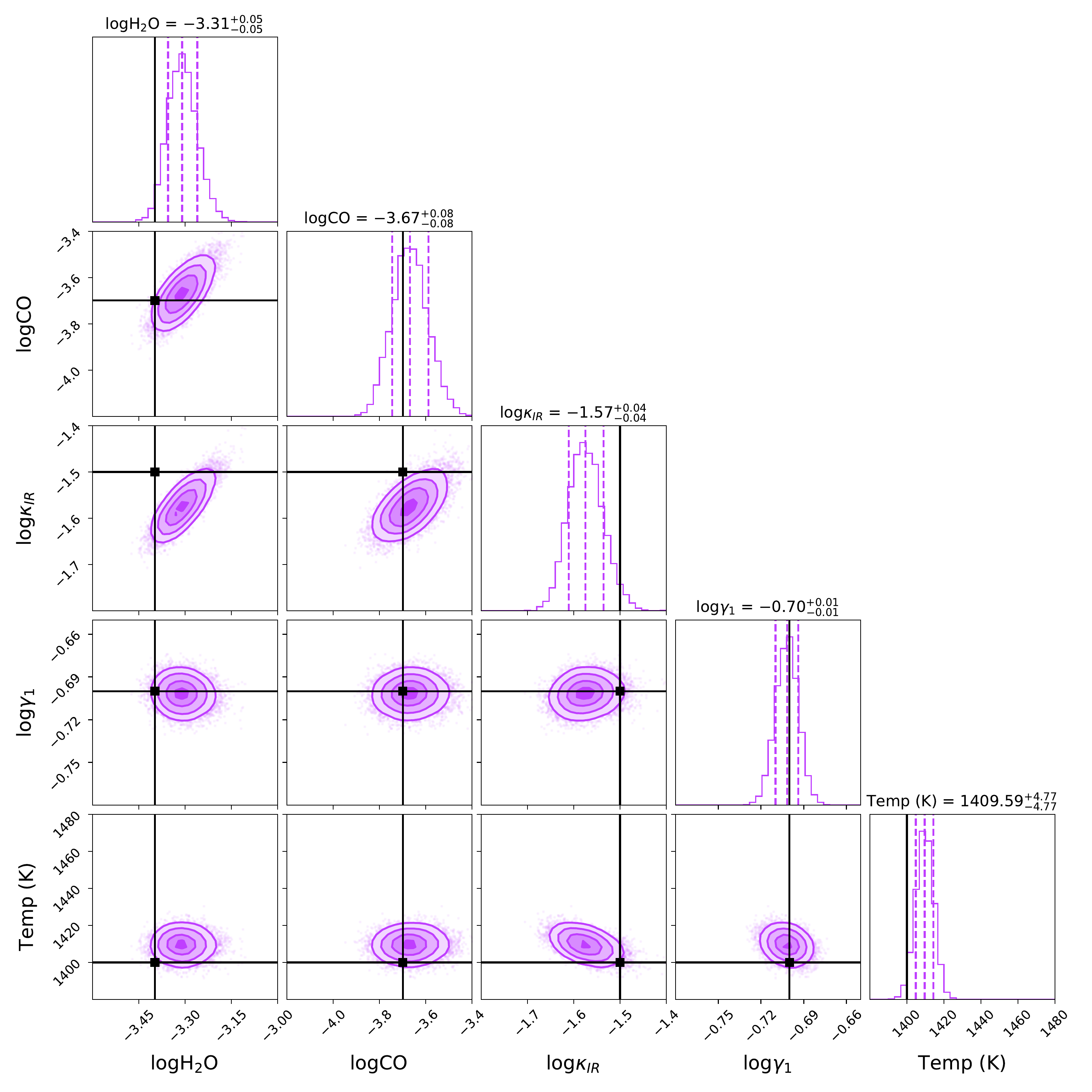}
    \caption{The posterior distribution for the retrieval benchmark test presented in Fig. \ref{fig:chivnem_spec}. The vertical black lines represent the input values used in the model that was generated using \textsc{CHIMERA}.}
    \label{fig:chivnem_triangle}
\end{figure*}

% Don't change these lines
\bsp	% typesetting comment
\label{lastpage}
\end{document}